\documentclass[11pt,a4paper]{article}
\usepackage{comment}
\usepackage{appendix}
\usepackage{amsfonts}
\usepackage{caption}
\usepackage{float}
\usepackage{subcaption}
\usepackage{graphicx}
\usepackage{mathtools}
\usepackage{amssymb}
\usepackage{amsmath}
\usepackage{amsbsy}
\usepackage{color}
\usepackage{dsfont}
\usepackage[dvipsnames]{xcolor}
\usepackage{upgreek}
\usepackage{soul}
\usepackage[normalem]{ulem}
\newcommand{\ee}{\end{equation}}
\newcommand{\bb}{\begin{equation}}
\newcommand{\be}{\begin{equation}}
\newcommand{\eqb}{\begin{eqnarray}}
\newcommand{\eqf}{\end{eqnarray}}

\definecolor{gre}{rgb}{0,0.4,0.3}
\usepackage{chngpage}

\usepackage[colorlinks=true,linkcolor=blue,urlcolor=cyan,citecolor = red]{hyperref}

\usepackage{appendix}

\newcommand{\Z}{\mathbb{Z}}
\usepackage{color}

\newcommand{\dr}[1]{\frac{d#1}{dr}}
\newcommand{\rdr}[1]{r \frac{d}{ d\,r}\left(\frac{#1}{r}\right)}

\def\a{\alpha}
\def\b{\beta}

\def\d{\delta}
\def\f{\phi}               

\def\g{\gamma}
\def\h{\eta}

\def\k{\kappa}            
\def\l{\lambda}
\def\m{\mu}
\def\n{\nu}

\def\s{\sigma}

\def\x{\xi}
\def\z{\zeta}

\def\dr{\frac{d}{dr}}

\usepackage{geometry}
\begin{document}
\begin{center}
\vspace*{2.5cm}
{\Large{\textbf{
Vortex solutions in the presence of Dark Portals
}}}

\vspace*{1.2cm}

{\large
Paola Arias$^{a}$,\,%
Ariel Arza$^{b}$,\,%
Fidel A. Schaposnik$^{c}$\footnote{Also at CICBA},\,%
Diego Vargas-Arancibia$^{a}$,\,%
Moira Venegas$^{a}$%
\\[3mm]
{\it{
$^{a}$ Departamento de Fisica, Universidad de Santiago de Chile, Casilla 307, Santiago, Chile
\\
$^{b}$ 
Institute for Theoretical and Mathematical Physics, Lomonosov Moscow State University (ITMP), 119991 Moscow, Russia\\
$^{c}$ Departamento de Fisica,Universidad Nacional de La Plata\\ Instituto de Fisica La Plata, CONICET\\C.C. 67, 1900 La Plata, Argentina.\\
}}}
\end{center}
\vspace*{1cm}

\begin{abstract}
  
 {The existence of hidden sectors weakly coupled to the visible one has been extensively studied as a way to extend the Standard Model (SM) and to provide a good dark matter candidate. In this work we analyze two models in which gauge and scalar hidden fields interact with the visible sector which, for simplicity we take as an Abelian Higgs model. In one of them, the hidden sector consists of an uncharged scalar. The connection with the visible Abelian Higgs sector is in this case provided by  the interaction between the scalars in the two sectors and also by the coupling  between the hidden scalar and the visible squared field strength. This model can be seen as an extension of previously studied ones  both in high energy and in condensed matter physics and the solutions that we find indicate that it could be relevant in connection to establishing the  dark matter relic density. In the second model, both sectors correspond to spontaneously broken gauge theories and hence they feature  classical vortex solutions.  Introducing three different portals between the two sectors, we analyze the dependence of  the model dynamics on the portal parameters and discuss the physical implications in connection with dark matter and gravitational radiation issues.} 
\end{abstract}
\newpage
\section{Introduction}
One of the major issues in physics nowadays is  how to extend the Standard Model (SM) such that successfully accounts for some missing pieces (neutrinos \cite{Gonzalez-Garcia:2002bkq}, strong CP \cite{Peccei:2006as}, among others) and also provides a solid dark matter candidate. From the particle physics point of view, effective field theories have been proposed in which  one or more particles - scalars, vectors and/or fermions -  from a hidden sector could emerge, potentially coupling to the visible sector through  ``portals",  interaction terms that open a channel between visible and hidden sectors that could leave   imprints on relevant observables. Portals via the Higgs particle are very popular, because it allow for  direct, renormalizable connection with the hidden sector. Such particles could then provide a stable candidate for dark matter.

Extensions of an Abelian-Higgs model have been largely considered  as  toy models  in order to build theories beyond the Standard Model (SM) in the low energy limit. The U(1) Abelian-Higgs stands as a toy model for SM-like theories with a gauge group $SU(2)\times U(1)$. As it is well-known,  such a model  supports classical vortices  \cite{Nielsen:1973cs} with very precisely known solutions \cite{deVega:1976xbp}. {Some examples of these kind of models are semi-local  and electroweak vortices \cite{Vachaspati:1991dz,Achucarro:1999it,hindmarsh1995cosmic}.} Thus, it seems relevant to analyze whether such configurations still hold and under what circumstances once the Abelian-Higgs model is modified by the inclusion of portal terms. Moreover, Abelian-Higgs vortices could  play an important role as dark strings.  Indeed, cosmic string networks  are hypothetical one dimensional  topological defects that could be formed during phase transitions in the early universe \cite{Kibble:1980mv,kolb2018early,vilenkin1994cosmic,hindmarsh1995cosmic,copeland2010cosmic,sakellariadou2005note}. Then, string radiation might lead to a sizable population of non-thermal dark matter  \cite{Brandenberger:1986vj, Vincent:1997cx, Matsunami:2019fss}. This mechanism has been particularly  studied in connection with the existence of axions \cite{Harari:1987ht, Hagmann:2000ja,Wantz:2009mi, Hiramatsu:2010yu, Kawasaki:2014sqa,  ramberg2019probing}, WIMPs \cite{Jeannerot:1999yn, Cui:2008bd} and hidden photons \cite{Long:2019lwl}.  

Different portal connecting the visible and hidden sectors have been considered 
\begin{itemize}
   \item \textbf{Kinetic mixing}: the kinetic mixing (KM) portal is a natural coupling to be  expected when  a second $U(1)$ symmetry is added, since it is gauge invariant and renormalizable. Originally introduced by \cite{okun1982limits, holdom1986searching}, the extra U(1) field - usually dubbed hidden or dark photon -  can be a   generic feature arising from string compactifications \cite{Abel:2008ai, Goodsell:2009xc, Cicoli:2011yh}. It connects the hidden photon with the hypercharge, but if the hidden field is lightweight, the dominant interaction is with the photon through a kinetic mixing term of the form $\xi F_{\mu\nu} G^{\mu\nu}$, with $F_{\mu\nu}$ the visible field strength, $G^{\mu\nu}$ the hidden sector one and  $\xi$ a dimensionless parameter, expected to be small.  It should be also noted that the supersymmetric extension require both gauge mixing and Higgs portal couplings and hence it is a necessary condition in order to have self-dual equations, see \cite{Arias:2014yua}.
   \item \textbf{Higgs portal}: another   renormalizable and gauge invariant interaction was   first discussed in \cite{patt2006higgs} in a proposal of possible extensions of the SM. It connects scalar fields $\phi$ (visible sector) and $\eta$ (hidden sector) via a term $\lambda |\phi^2 \|\eta|^2$, with $\lambda$ a dimensionless coupling.  Such interaction it is widely known as Higgs portal (HP) because it has been heavily exploited identifying $\phi$ as the Higgs field (see for instance \cite{Arcadi:2019lka} and references therein). 
   \item \textbf{Scalar-gauge effective interactions:} corresponds to non-renormalizable terms with mass dimension $\text{dim}>5$ mediated by a  mass scale $\Lambda$. An example of this scalar-gauge interaction (SGI portal) is
\begin{eqnarray}
\mathcal{L}_{\Lambda}=\frac{g^2}{\Lambda^2}|\f|^2G_{\m\n} G^{\m\n},
\end{eqnarray} 
where $g$ is the electric charge in the hidden sector.
A possible way to obtain the SGI term is to include quantum corrections,  for instance, originated through one-loop like vector-triangle or heavy charged fermions
 \cite{Djouadi:2005gi,Carena:2012xa,low2011singlet}. {In this context, $\Lambda$ can be expressed in terms of the charge $g$ and HP coupling parameter $\l$}.
 On the other hand, a more pragmatic approach is to consider the SGI as an effective portal regardless of its physical origin \cite{sato2011scalar}. {In this article we have considered the latter alternative where $\Lambda$ is understood as a free parameter.} 
\end{itemize}
{As previously remarked,} in this work we shall consider  for simplicity, a visible sector  with   a  $U(1)$   spontaneously broken  gauge theory coupled to a complex scalar. Regarding  the hidden sector we shall also consider a $U(1)$ gauge theory but discuss its coupling to either a complex or a real scalar. As for the coupling of the two sectors we shall investigate models with  combination of the  portals described above,  with particular interest in the the role of the SGI one whose effects have not been previously studied in detail.

In section \ref{sectionunbroken} the Lagrangian governing the hidden field dynamics  just consists of a Maxwell term plus a real scalar field so that there is no gauge symmetry breaking in this sector. Nevertheless, the mixing Lagrangian, apart from   KM and SGI interactions includes a potential mixing between visible and hidden scalar fields in such a way that the latter  develops a non-zero value at the origin, which can be associated with a scalar condensate, a phenomenon of interest already studied in the context of cosmic strings \cite{Witten:1984eb} and also topological defects \cite{2002, shifman2013simple, PerezIpina:2018jcg}. Additionally,  is particularly relevant in condensed matter physics.

The case in which both the visible and hidden sectors correspond to spontaneously broken gauge theories coupled to a complex scalar is discussed in section \ref{broken}. The symmetry breaking potentials are chosen to be those required in order to have first order Bogomolny equations when just the HP and KM portals are present except that in our case we leave the three coupling constants as independent parameters \cite{Arias:2014yua} but  the addition of a  SGI portal requires consider the second order Euler-Lagrange equations. An axially symmetric ansatz \`a la Nielsen-Olesen leads to a system of coupled non-linear equations which we solve numerically.

In section \ref{sec:conclus} we comment on our results and we discuss possible  physical implications of the models in connection with dark matter and gravitational radiation issues.

\section{A model with an uncharged scalar} \label{sectionunbroken}
{\subsection{The model}}

We start by considering a visible sector - composed of an Abelian Maxwell-Higgs theory with a $U_A(1)$ gauge field $A_\mu$ and a charged scalar $\phi$,  interacting with a hidden sector which, by completness is composed of a $U_G(1)$ gauge field $G_\mu$ and a {\it real} scalar $\eta$, via three kind of portals: a Higgs portal (HP), kinetic mixing term (KM) and an effective interaction (SGI) between the hidden scalar and the visible field strength, given by
\begin{eqnarray}
\mathcal{L}_{\Lambda}=-\frac{e^2}{\Lambda_e^2} F_{\mu\nu} F^{\mu\nu}|\h|^2.
\end{eqnarray}
Here $F_{\mu\nu}$ is the visible field strength, $e$ is the visible electric charge and $\eta$ the scalar in the hidden sector. 
As mentioned in the introduction, variations of this coupling linking scalars with gauge bosons have been used in different contexts such as dark radiation in early times \cite{Jeong:2013eza} or in physics of the Higgs mechanism \cite{2021,Dawson:2018dcd}. It can be induced by loop corrections, but in our case we will follow the approach of \cite{sato2011scalar} and treat $\Lambda_e$ as a given high energy scale. 

The coupling between the scalars from the two sectors is given by
\be
V= \gamma\left[\left(-\tilde \mu^2+|\phi|^2\right)\eta^2+ \zeta \eta^4\right]+\frac{\lambda}4 (|\phi|^2 - v^2)^2.\label{UBpotential}
\ee
We shall then consider the  Lagrangian density of our model, as the sum of the visible and hidden sectors, plus the mixing interactions between them 
\be
\mathcal L= \mathcal L_{mix}+\mathcal L_{HS}+\mathcal L_{visible},\label{UNlag1}
\ee
where
 \begin{eqnarray}
\mathcal{L}_{visible}&=& -\frac14 F_{\mu\nu} F^{\mu\nu} - \frac12|D_\mu^{(A)}\phi|^2 - \frac{\lambda}4 (|\phi|^2 - v^2)^2, \nonumber  \\
\mathcal{L}_{HS} &=& -\frac14 G_{\mu\nu} G^{\mu\nu}-\frac{1}2\left(\partial_\mu\eta\right)^2,
\nonumber\\
\mathcal{L}_{mix}& =& \frac{\xi}2 G_{\mu\nu}F^{\mu\nu} -{{ \frac{e^2}{\Lambda^2_e}\,\h^2 F_{\mu\nu}F^{\mu\nu} }}-\gamma\left[\left(-\tilde{\mu}^2+|\phi|^2\right)\eta^2+\zeta \eta^4\right].
\label{eq:s3_Lag}
\end{eqnarray}
Here the covariant derivatives are defined as
\be
D_\mu^{(A)}=\partial_\mu-ie A_\mu
\ee 
Note that at large distances, where $A_\mu$ and $G_\mu$ fields strength should vanish, the $\eta$ potential in ${\cal L}_{mix}$ vanishes for $\eta = 0$.

{The model with just 3 parameters in the hidden sector, $\gamma, \zeta$ and $\tilde{\mu}$, has been used in the literature in the context of SM extensions, identifying the visible scalar field $\phi$  
 with the Higgs field and $\eta$ with a dark matter scalar. By requiring the energy density to account for the observed dark matter relic abundance  and  successfully satisfy the existing observational constraints, it is possible to find the viable values for these parameters \cite{burgess2001minimal,Barger:2007im}.} 

In our search of vortex solutions for this model we are specially interested in the emergence of a hidden scalar condensate at $r=0$, triggered by an energy decrease leading to a stable condensate configuration  at expenses of the   visible scalar energy. This ``competing order'' phenomenon   has been previously discussed in the context of superconducting superstrings \cite{Witten:1984eb}, topological defects \cite{shifman2013simple,PerezIpina:2018jcg}  and, remarkably,  in the context of condensed matter physics concerning the study of high-temperature superconductors \cite{2002}. 

{We shall consider the potential  $V$ to be  bounded from below and determine  the conditions on its parameters in connection with the symmetry-breaking pattern.} 
{A study of the extrema of the potential energy in eq.~(\ref{eq:s3_Lag}) shows that there is a stable local minimum at $r\rightarrow \infty$ given by  ($\langle \phi\rangle=v$, $\langle \eta\rangle=0$),
 if the following condition is satisfied
\be
2\gamma(v^2-\tilde \mu^2)>0,\,\,\,\, \Rightarrow v>\tilde \mu.
\label{tresdos}
\ee}
A second stable minimum at $r\rightarrow \infty $ is possible given by  ($\langle \phi\rangle$=0, $\langle \eta\rangle=\tilde{\mu}/\sqrt{2\zeta}$), if the following condition is satisfied
\bb
\tilde{\mu}^2>\frac{\zeta \lambda v^2}{\gamma}.
\ee
Both local minima coexists as long as $\gamma>\lambda \zeta$ in which case, the extremum ($\langle \phi\rangle=v$, $\langle \eta\rangle=0$) is deeper whenever the relation
\be
\tilde{\mu}^2<v^2\sqrt{\frac{\lambda \zeta}{\gamma}},
\label{eq:s3_condicion2_minimo}
\ee
holds. We shall consider in the following that the expectation value of the fields at $r\rightarrow \infty$ is given by  ($\langle \phi\rangle=v$, $\langle \eta\rangle=0$) and therefore the relations of eqs.~(\ref{tresdos}) and (\ref{eq:s3_condicion2_minimo}) hold.

It is possible to diagonalize the kinetic part of the gauge fields by performing a shift of the hidden gauge field $G_\mu$:
\be
G_\mu \rightarrow G_\m+\x A_\m, 
\label{3.7}
\ee
and redefine the visible gauge field  $ A_\m \rightarrow  {A_\m}/{\sqrt{1-\x^2}}$ with  $ e\rightarrow  {e}/{\sqrt{1-\x^2}}$. In the present case, this shift  decouples $G_\mu$ from the field equations and it magnetic field turns out to be given by  
\be B_G=\xi B_A.
\label{Bes}
\ee
This is analogue to what has been found in ref.~\cite{arias2014vortex}. After  the shift \eqref{3.7} the  kinetic mixing portal is absent from the Lagrangian which is left just with the scalar mixing term  (HP) and the  SGI interaction\footnote{We have not considered a Higgs-dark vector field   portal because it cannot be  induced since $\eta$ is uncharged under the $U_G(1)$.} Therefore, we will not include the hidden magnetic field in our discussions from now on, since it will just follow the behaviour of the visible magnetic field, diminished by $\xi$. This leaves a simpler problem to solve.

In order to find solutions to the field equations associated to  Lagrangian density \eqref{UNlag1} it  will be convenient to redefine variables and fields
\bb
r\rightarrow \frac{r}{(e v)}, \,\,\,\,\, A\rightarrow A v, \,\,\,\,\, G\rightarrow G v,\,\,\,\,\, \phi\rightarrow \phi v, \,\,\,\,\, \eta\rightarrow \eta v,
\ee
 so that that associated  (full) Lagrangian reads
\begin{eqnarray}
&& L=\frac{v}e \int d^3x\,\left(\tilde{ \mathcal{L}}_{visible}+\tilde{\mathcal{L}}_{HS} +\tilde{\mathcal{L}}_{mix}\right)\nonumber \\
&&\tilde{\mathcal{L}}_{visible}=-\frac{1}4 F_{\m\n}F^{\m\n}-\frac{1}{2}|(\partial_\m-i A)\f|^2-\frac{\k}{4}(|\f|^2-1)^2 \nonumber \\&&\tilde{\mathcal{L}}_{HS}=-\frac{1}4G_{\m\n}G^{\m\n}-\frac{1}{2} \partial_\m \h \partial^\m \h \nonumber \\
&&\tilde{\mathcal{L}}_{mix}= -\Phi_e^2 \h^2 F_{\m\n}F^{\m\n}-\Gamma\left[\left(-{\mu}^2+|\phi|^2\right)|\eta|^2+\zeta |\eta|^4\right],
\label{eq:s3_Lagtilde}
\end{eqnarray}
where   \be\Phi_e^2=\frac{e^2\,v^2}{\Lambda_e^2}, \quad \Gamma\equiv\frac{\gamma}{e^2}, \quad \mu\equiv \frac{\tilde \mu}{v}, \quad \k\equiv\frac{\l}{e^2}.
\ee
Notice that because of gauge symmetry breaking, not only the visible sector fields $\phi$ and $A_\mu$ acquire masses, but also $\eta$ becomes a massive real scalar  
\be
\frac{m_\phi}{ev}=\sqrt{2\kappa} \;, \,\,\,\,\,\,  \frac{m_A}{ev}=1\;, \,\,\,\,\,\, \; \frac{m_\eta}{ev}=\sqrt{2\Gamma(1-\mu^2)}\;.
\label{eq:s3_masses}
\ee
The mass of $\eta$ is consistent with the condition $\mu<1$ in \eqref{tresdos}. 

Static finite energy solutions of the field equations require  
$A_0=0$ \cite{Manias:1985nr}. Moreover,  since we are interested in finding  vortex solutions in the visible sector, we will  look for   $z$-independent field configurations proposing the well-honored  Nielsen-Olesen \cite{Nielsen:1973cs}  axially symmetric ansatz for fields,    namely
\be
A_\varphi=n\frac{\alpha (r)}r \;, \ \ \ \   \phi= f(r)e^{in\varphi} \;,
\ee
with $n$ an integer. For the field $\eta$, the only hidden sector field involved in our analysis,  we take
\be
\eta=h(r).
\ee 
The energy per unit length then becomes
\eqb
    \frac{E}{\ell} = &&\int d^2 r\frac{n^2}{2r^2}\left(\frac{d\alpha}{dr} \right)^2+\frac{1}{2}\left( \frac{d f }{dr}\right)^2+\frac{1}{2}\frac{n^2}{r^2}(1-\alpha)^2f^2+\frac{\kappa}{4}(f^2-1)^2+\frac{1}{2}\left(\frac{dh}{dr} \right)^2  \nonumber \\
    && +2\Phi_e^2\frac{n^2}{r^2}\left(\frac{d \alpha}{dr}\right)^2 h^2+ \Gamma\left[(-\mu^2+f^2)h^2+\zeta h^4\right]\equiv \int \varepsilon\, d^2r. \label{eq:s3_energy}
\eqf 

The boundary conditions at $r=0$ and $r\rightarrow\infty$ for the visible sector fields which is compatible with a well defined, finite energy, are 
\eqb
f(0)=\alpha(0)=0,\ \ \ \ \  f(\infty)=\alpha(\infty)=1.
\eqf
Concerning the radial visible scalar function  $h(r)$, it has to vanish asymptotically as $r\rightarrow\infty$,  a   condition     that ensures gauge symmetry breaking in the visible sector. On the other hand, at $r=0$  there is no a priori condition for $h(r)$. Now,  performing a  Frobenius expansion  \cite{PerezIpina:2018jcg}, it can be checked  that  in order to preserve a non zero value of $h$ at the origin and hence a competing order, its first derivative  should vanish at the origin. Therefore the boundary conditions for the hidden sector scalar reads
\eqb
h'(0)=h(\infty)=0.
\eqf

Where $h'(r)$ stands for $dh/dr$. After the axially symmetric ansatz (for simplicity with $n=1$)  the equations of motion associated to Lagrangian \eqref{eq:s3_Lagtilde} become
\begin{eqnarray}
&&\dr\left[\left(1+4\Phi^2_e h^2\right)\frac{\a'}{r}\right]+\frac{f^2}{r}(1-\a)=0  \label{eq:aunbroken}  \\ 
&& \frac{1}{r}\dr \left(r f'\right)-\frac{f}{r^2}(1-\a)^2 -\k(f^2-1)f-2\Gamma f h^2=0 \label{eq:funbroken} \\
&& \frac{1}{r}\dr \left(r h'\right)-2\Gamma h(f^2-{\m}^2)-4\Gamma \z h^3-4  \Phi^2_e \frac{\a'^2}{r^2}h=0.\label{eq:hunbroken}
\end{eqnarray} 
{ We shall analyze below the solution of these equations both numerically and also studied analytically  the fields behavior for  $r\rightarrow 0$ and $r \rightarrow \infty$.} 

\begin{figure}[t!]
	\centering
	\begin{subfigure}[b]{0.45\textwidth}
	\centering
   \includegraphics[width=1\textwidth]{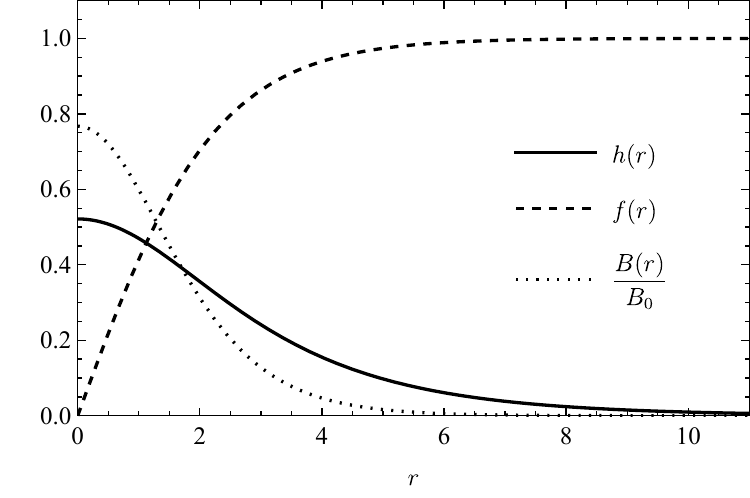}
	\caption{}
	\label{fig:s3_profile_energy_p}
	\end{subfigure}%
	 \hfill
	\begin{subfigure}[b]{0.45\textwidth}
	\centering
    \includegraphics[width=1\textwidth ]{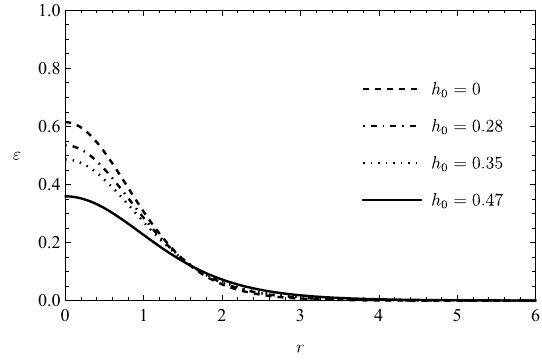}
	\caption{}
	\label{fig:s3_profile_energy_e}
	\end{subfigure}%
    \caption{(a) Visible and hidden fields profile as a function of the distance from the center of the core for $\Gamma=0.6$, $\zeta=0.85$, $\mu=0.91$, $\kappa=0.5$ and the SGI portal  turned off ($\Phi_e=0$). (b) Energy density as a function of the distance from the core for several values of $h_0$. To obtain the values of $h_0$ we have set $\zeta=0.85$, $\mu=0.91$ and $\kappa=0.5$ and varying $\Gamma$. 
    }
    \label{fig:s3_profile_energy}
\end{figure}
\subsection{Numerical results and analytical estimates}
 Figure \ref{fig:s3_profile_energy} shows  the numerical solution of the system of equations \eqref{eq:aunbroken}-\eqref{eq:hunbroken} using a shooting method.  In fig.~\ref{fig:s3_profile_energy_p} we  show the profiles of  $f(r)$, $h(r)$ and the magnetic field, normalized by   the magnetic field in the absence of any portal ($\Gamma=\Phi_e=0$),  as a function of the distance $r$ from the core. The potential parameters  were chosen following eqs.~(\ref{tresdos}) and (\ref{eq:s3_condicion2_minimo}), such that $\eta$ displays   a condensate around the origin. The strength of the couplings chosen to present our results will be commented  at the end of the section. As a consequence there is a decrease of the kinetic energy of the visible scalar, delaying the advent of $f(r)=1$  to a larger $r$ value. The existence of the hidden scalar condensate at the origin is due to an energetically favorable condition for the system, as it is also supported by fig.~\ref{fig:s3_profile_energy_e} where the total energy density eq.~(\ref{eq:s3_energy}) is shown. In this figure we have denoted $h_0\equiv h(0)$ and the curves correspond to different configurations with $h_0=0$ in the case with no condensate . As we can see, the larger the condensate energy ({\it i.e.} a larger $h_0$) the less the total energy density due to the fact that the potential energy associated to   $\eta$  and the   kinetic energy associated to $\phi$ decrease near the origin when the condensate sets up. 

\begin{figure}[t!]
	\centering
	\begin{subfigure}[b]{0.45\textwidth}
	\centering
   \includegraphics[width=1\textwidth]{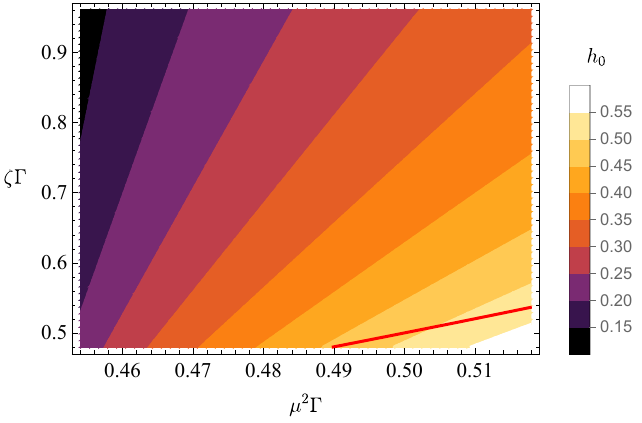}
	\caption{\label{fig:h0_zg}}
	\end{subfigure}%
	 \hfill
	\begin{subfigure}[b]{0.45\textwidth}
	\centering
    \includegraphics[width=1\textwidth ]{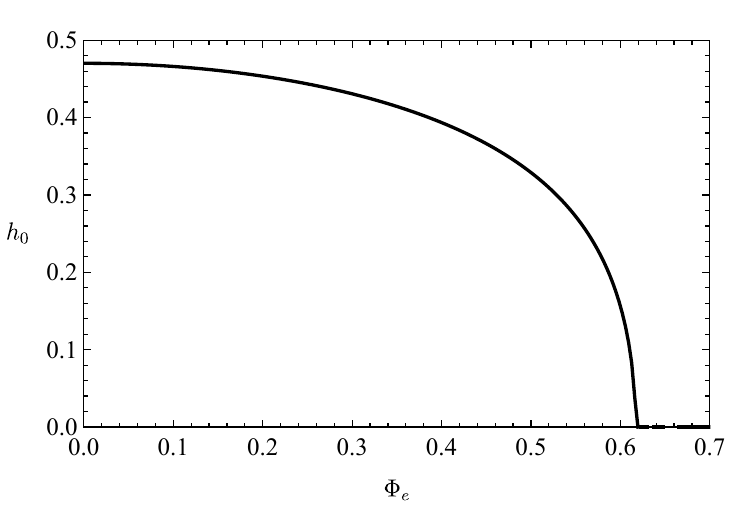}
	\caption{\label{fig:s3_h0Phi}}
	\end{subfigure}%
    \caption{(a) The amplitude values  $h(0)\equiv h_0$ as a function of the parameters $\Gamma \mu^2$ and $\Gamma \zeta$ considering $\Gamma=0.6$. Above the red line it is  satisfied that $\langle \eta \rangle = 0 $, $\langle \phi\rangle=v$ is the global minimum of our model. We have considered $\kappa$= 0.5 and $\Phi_e$= 0. (b) The amplitude values  $h_0$ as a function of the parameter $\Phi_e$. The rest of parameters remain as $\kappa$= 0.5, $\zeta$= 0.85, $\mu$= 0.91 and to make sure that the minimum corresponds to the global minimum we have taken $\Gamma=0.6$.  }
    \label{fig:s3_plotsh0}
\end{figure}

 {We show in fig.~\ref{fig:h0_zg} the values of $h_0$ as a function of  $\Gamma \mu^2$ and $\Gamma \zeta$. We can see from the figure that increasing $\mu^2 \Gamma$ -  the $\eta^2$ coupling -  increases the condensate energy at the origin, while $\zeta \Gamma$ - the  $\eta^4 $ coupling  -  remains small. This is due to the fact that   $\mu^2\Gamma$ gives a negative contribution to the energy of the hidden scalar, while $\zeta\Gamma$ gives a positive one. The region above the red line   corresponds to the parameter space where  ($\langle \phi\rangle=v$, $\langle \eta\rangle=0$) is the global minimum of the system. } In fig. \ref{fig:s3_h0Phi} we show the change in $h_0$ as a function of the SGI parameter, $\Phi_e$, with the rest of the parameters fixed (see the caption for more details). For values of $\Phi_e$ bigger than some threshold (for this configuration at  $\Phi_e=0.62$), the condensate vanishes.

In order to have a better understanding {of the behavior described above}, we have solved the coupled equations \eqref{eq:aunbroken}-\eqref{eq:hunbroken} analytically in the limits of $r\rightarrow 0$ and $r\rightarrow \infty$. A detailed derivation is left for the Appendix \ref{app:one_unbroken},  in here we just summarize the most important findings. 

\subsubsection*{Analytical behaviour near zero and infinity}
Concerning the scalar fields, we find the following behaviour near  the origin that is in correspondence with the numerical solution
\begin{eqnarray}
&&f(r) \approx \frac{f_1 \,}{2}\,r, \\
&&h(r)\approx h_0\left(1-\frac{Q}{4}r^2\right).
\label{eq:s3_hzero}
\end{eqnarray}
 {with $f_1$ an integration constant. The growth of $f(r)$ is controlled by $U_f$  }
\bb
U_f^2=\k-2\Gamma h_0^2.
\label{eq:s3_Uf}
\ee
so that  in order to guarantee  spontaneous symmetry breaking, $U_f^2$ should be positive definite. We have checked that this is so  when condensation takes place at the origin. Concerning   the visible scalar kinetic energy in this case it takes a lower value.

On the other hand, for the hidden scalar, we have that in order for the analytical expression given by equation \eqref{eq:s3_hzero} to match the numerical solution shown in fig.~\ref{fig:s3_profile_energy_p}, the coefficient $Q$ has to be positively defined, with  
\bb
Q=2\Gamma \m^2-4\Gamma \z h_0^2-4\Phi_e^2B_0^2 ,
\ee
where $B_0$ is the value of the visible magnetic field at  the origin. The requirement $Q>0$ imposes the following condition
\bb
\frac{ \mu^2}{2\zeta}> h_0^2+\frac{\Phi_e^2 B_0^2}{\zeta\Gamma}.
\ee
{This relation imposes an upper limit for the amplitude of $h(r)$ implying the existence of a threshold $\Phi_e$ for the condensate to exist, see   fig.~\ref{fig:s3_h0Phi}.} In the absence of the SGI portal, the amplitude $h_0$ will never reach the value $\mu/\sqrt{2 {\zeta}}$ which, together with $\langle \phi\rangle=0$, are the expectation values of the fields, {this implying that there is  no symmetry breaking in the visible sector.}

From our previous discussion we conclude that the existence of a $\eta$ condensate makes the correlation length of $\phi$  larger, postponing the establishment  of $|\phi|\rightarrow 1$  to a larger $r$. {Also,  the condensate existence requires that the hidden scalar tends to zero before the visible scalar reaches its vev. This takes place provided the following inequalities hold}
\bb
Q^{-1}\ll U_f^{-2} \Rightarrow 2\Gamma\mu^2-4\Phi_e^2B_0^2\gg \kappa.
\ee
{One can check in   fig.~(\ref{fig:h0_zg})  that the whole parameter range above the red line where the condensate can emerge satisfies the above condition.}

Concerning the  short distant behavior for the  vector field function $\alpha(r)$ leads to a magnetic field given by
\be
B(r)\sim const(1-4\Phi_e^2 h^2).
\ee
which shows that the presence of the condensate reduces the   the visible magnetic field amplitude.

Since at $r
\rightarrow \infty$ there is no  condensate and the SGI interaction vanishes, the $f(r)$ and $\alpha(r)$ asymptotic behaviour   corresponds to the case of a  Maxwell-Higgs model (see Appendix \ref{app:one_unbroken}) 
\begin{eqnarray}
&&\a(r)=1+\a_1\,r K_1(r),\\ 
&& f(r)=1+f_1 K_0\left(\frac{m_\f}{e v}r\right), \\
&& h(r)=h_1 K_0\left(\frac{m_\h}{e v}r\right).
\end{eqnarray} 
with  $\alpha_1, f_1 $ and $h_1$  integration constants. The hidden scalar decreases exponentially to its vacuum value, controlled by its mass.
\begin{figure}[t!]
     \centering
     \begin{subfigure}[b]{0.45\textwidth}
         \centering
         \includegraphics[width=\textwidth]{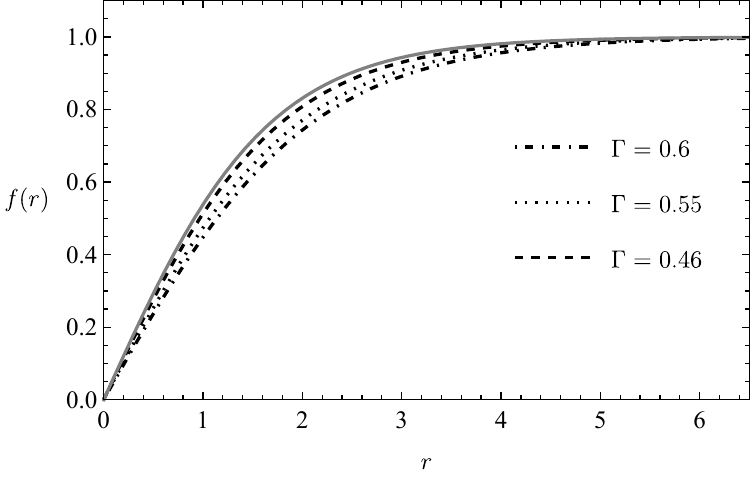}
         \caption{}
         \label{fig:f_G}
     \end{subfigure}
      \hfill
     \begin{subfigure}[b]{0.45\textwidth}
         \centering
         \includegraphics[width=\textwidth]{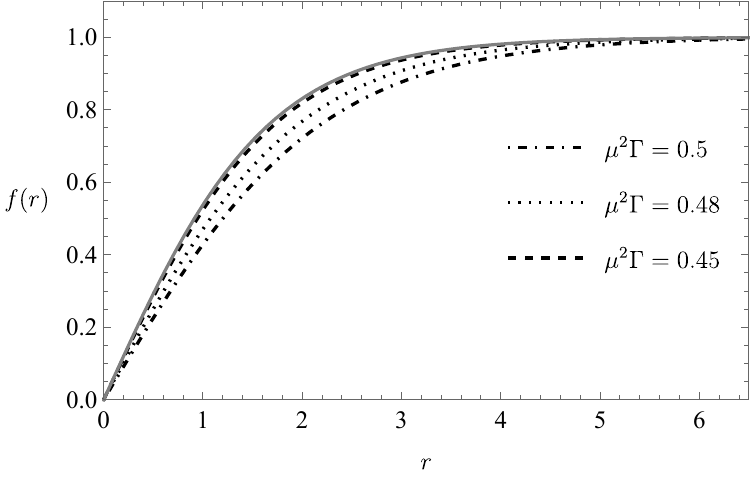}
         \caption{}
         \label{fig:s3_f_Gmu}
     \end{subfigure}

\bigskip
     \begin{subfigure}[b]{0.45\textwidth}
         \centering
         \includegraphics[width=\textwidth]{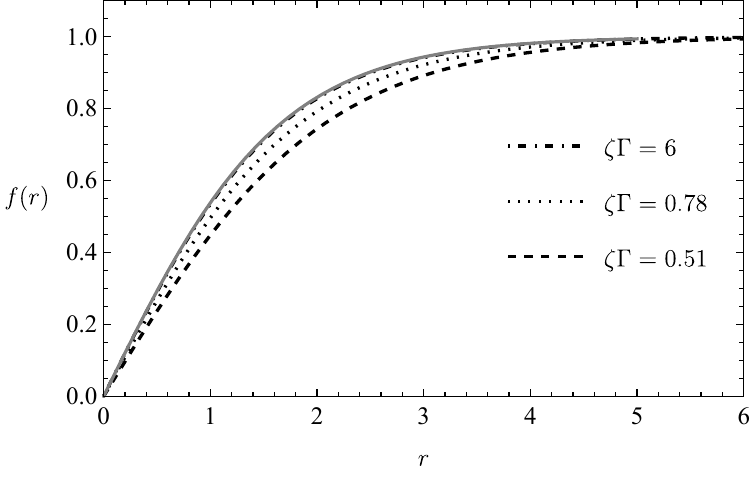}
         \caption{}
         \label{fig:s3_f_Gz}
     \end{subfigure}
          \hfill
     \begin{subfigure}[b]{0.45\textwidth}
         \centering
         \includegraphics[width=\textwidth]{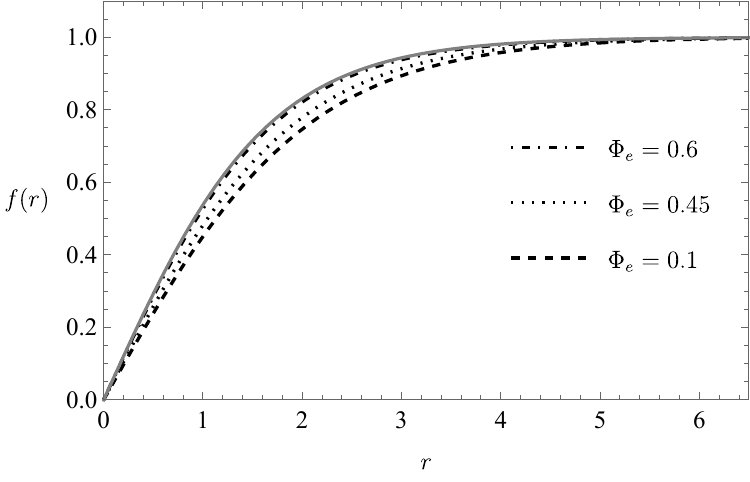}
         \caption{}
         \label{fig:f_Phi}
     \end{subfigure}
\caption{Visible scalar field profiles for different  values of the parameter of each portal. (a) Variation of $\Gamma$, we have fixed the rest of parameters as $\zeta=0.85$, $\mu=0.91$, $\kappa=0.5$ and $\Phi_e=0$; (b) Variation of $\Gamma \mu^2$, we have fixed $\zeta=0.85$, $\kappa=0.5$ and $\Phi_e=0$; (c) Variation of $\Gamma \zeta$, we have fixed $\mu=0.91$, $\kappa=0.5$ and $\Phi_e=0$; finally, (d) variation of $\Phi_e$, where the rest of parameters correspond to $\zeta=0.85$, $\Gamma=0.6$, $\mu=0.91$ and $\kappa=0.5$. In the four plots we have set the parameters $v$,$e$ equal to unity. }
\label{fig:s3_f}
\end{figure}
\subsection*{Numerical results}
We now  describe the results of the numerical solution of eqs.  \eqref{eq:aunbroken} to \eqref{eq:hunbroken}. We shall  first study the response of the visible fields to changes in the values of the parameters of the potential of the hidden sector, and the HP portal,  maintaining  the SGI portal off for a better understanding. Then, we will analyse the action of the SGI interaction.  

We show in fig.~\ref{fig:s3_f}  the behaviour of  the visible scalar  for different values of the HP portal parameter $\Gamma$, the couplings of the hidden potential and the SGI portal for the case in which condensation takes place. 
Concerning fig.~\ref{fig:f_G} in which the SGI interaction is off, we see  that   that a larger HP coupling $\Gamma$ increases the correlation length of $\phi$, as is to be expected from  eq.~(\ref{eq:s3_Uf}). The gray solid line corresponds to the scalar field in absence of any hidden sector coupling. As it is to be expected,  an increase in the  HP the coupling   leads to a larger condensate effect. In fig.~\ref{fig:s3_f_Gmu} we vary the  $\eta^2$ coupling, keeping  fixed the other parameters with  SGI portal still off. We can see that a larger coupling  (which corresponds to a larger   condensate amplitude, see fig.~\ref{fig:h0_zg}) deviates $f(r)$ from the solid gray line. An analogous behavior can be observed in  fig.~\ref{fig:s3_f_Gz}, where the profile of $f$ is presented for several value of the $\eta^4$ coupling, $\Gamma \zeta$.   The amplitude of the hidden condensate grows  for  smaller values  of the coupling, and this leads to a delay in the rise of the visible scalar field. In both cases this is due to the fact that the condensate growth takes place at expenses of the kinetic energy of   $\phi$. We note  that small changes in $\mu^2\Gamma$ (directly related to the hidden scalar mass, see eq.~(\ref{eq:s3_masses})) has an impact on the visible scalar behaviour.  

\begin{figure}[H]
     \centering
     \begin{subfigure}[h]{0.45\textwidth}
         \centering
         \includegraphics[width=\textwidth]{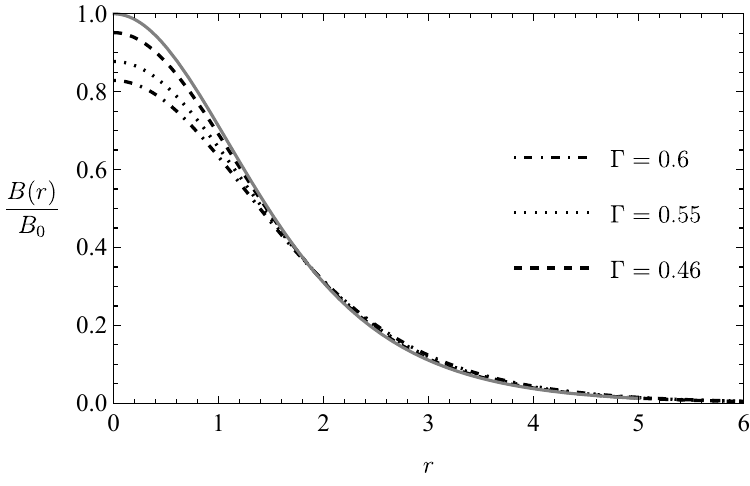}
            \caption{}
         \label{fig:B_G}
     \end{subfigure}
     \hfill
     \begin{subfigure}[h]{0.45\textwidth}
         \centering
         \includegraphics[width=\textwidth]{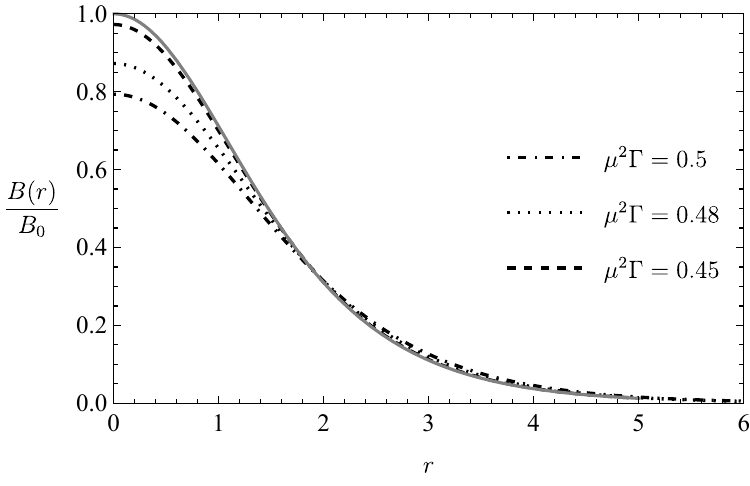}
          \caption{}
         \label{fig:B_Gmu}
     \end{subfigure}

\bigskip
     \begin{subfigure}[h]{0.45\textwidth}
         \centering
         \includegraphics[width=\textwidth]{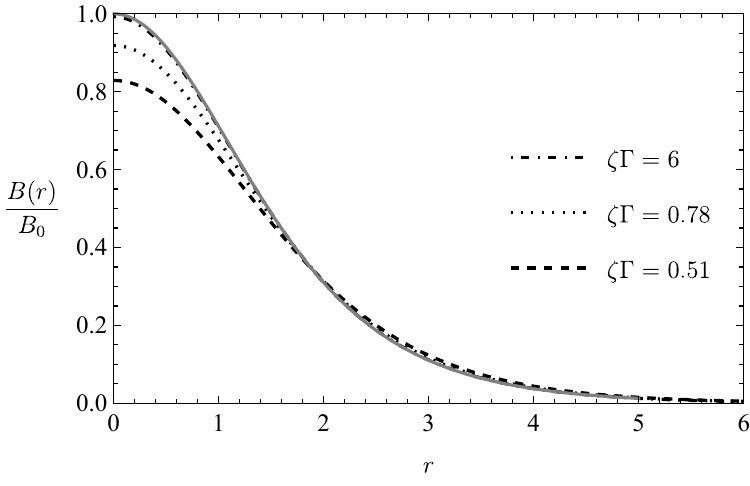}
          \caption{}
         \label{fig:B_Gz}
     \end{subfigure}
          \hfill
     \begin{subfigure}[h]{0.45\textwidth}
         \centering
         \includegraphics[width=\textwidth]{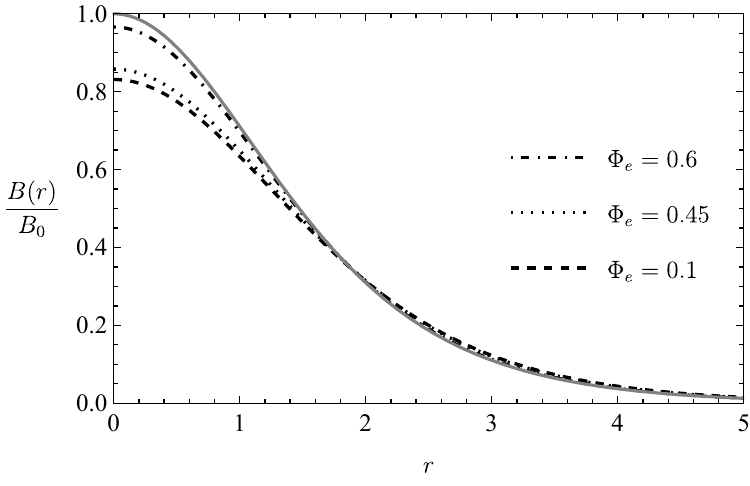}
          \caption{}
         \label{fig:B_Phi}
     \end{subfigure}
\caption{Visible magnetic field profiles for different  values of the parameter of each portal. (a) Variation of $\Gamma$, we have fixed the rest of parameters as $\zeta=0.85$, $\mu=0.91$, $\kappa=0.5$ and $\Phi_e=0$. (b) Variation of $\Gamma \mu^2$, we have fixed $\zeta=0.85$, $\kappa=0.5$ and $\Phi_e=0$. (c) Variation of $\Gamma \zeta$, we have fixed $\mu=0.91$, $\kappa=0.5$ and $\Phi_e=0$. Finally, (d) variation of $\Phi_e$, where the rest of parameters correspond to $\zeta=0.85$, $\Gamma=0.6$, $\mu=0.91$ and $\kappa=0.5$. In the four plots we have set the parameters $v$,$e$ equal to unity.}   
\label{fig:BUB}
\end{figure}

We display the profile of the visible magnetic field  in fig.~\ref{fig:BUB}. The plots   confirm  that whenever the parameters of the hidden scalar potential are chosen such that a larger condensate of $\eta$ at origin, the visible magnetic field is reduced, in accordance  with what has been found for the visible scalar. The effect of the SGI portal increases the visible magnetic energy reduces the amplitude and kinetic energy of the hidden condensate. Let us recall that in this model,  the hidden magnetic field con only arises if the KM portal is turned on and its behavior  is just  a re-scale  of the visible one so the above discussion applies for the hidden magnetic field.

Finally, let us comment on the possible phenomenology of the model. From the particle physics point of view, as we have already stressed, a very similar model (usually without the SGI interaction) has been extensively considered as an straightforward extension of the SM that can accommodate a viable DM candidate. It is usually assumed that the abundance of the DM is set by a freeze-out of Higgs-mediated interactions and the constraints are set on the value of the HP $\gamma$ and the DM mass $m_\eta$ given in Eq.~(\ref{eq:s3_masses}). For the typical benchmark values used here, we obtain a DM in a rather light mass region: $m_\eta\sim m_\phi/2$ combined with a HP coupling of $\gamma\sim 0.5$. Notice that even in the simplest scenario of standard cosmology freeze-out DM, such parameter space could account for a  fraction of the DM density (see for instance \cite{Barger:2007im,Cline:2013gha}). However, it is interesting  that the formation of a DM condensate as the one presented here could account for at least a fraction of the DM density in a non-thermal way. It is left as a future work to study if the condensate could be generated in the large mass region $m_\eta\gg m_\phi$ and/or in a weak-coupling regime.  

\section{A model with spontaneously broken gauge symmetry  in the visible and hidden sectors} \label{broken}
{\subsection{The model}}
In this section, we will discuss the interaction between a   Maxwell-Higgs visible sector with a  $U_A(1)$ gauge field $A_\mu$ and a charged scalar $\phi$ and a  hidden  sector with a $U_G(1)$ gauge field $G_\mu$ and a charged scalar $\eta$.   

We shall consider the three kind of interaction between the   two sectors, \textit{i.e.} a   kinetic mixing (KM) portal, a Higgs portal (HP) and an   effective interaction (SGI) between the visible scalar field and the hidden field strength, which for the model we study here will be of the form 
\begin{eqnarray}
\mathcal{L}_{\Lambda}=-\frac{g^2}{\Lambda_g^2} G_{\mu\nu} G^{\mu\nu}|\phi|^2.
\end{eqnarray}
In principle, we could have also considered the term proportional to $F_{\mu\nu} F^{\mu\nu} |\eta|^2$, nonetheless, seems redundant given that we are only interested in capturing the main effects of the interaction on the theory, so we do not include it. 

Putting all together, the corresponding Lagrangian densities of this model are
\eqb
\mathcal L_{visible} &=& -\frac14 F_{\mu\nu} F^{\mu\nu} - \frac12|D_\mu^{(A)}\phi|^2 - \frac{\lambda_1}4 (|\phi|^2 - v^2)^2, \label{2o}\\
\mathcal L_{HS} &=& -\frac14 G_{\mu\nu} G^{\mu\nu}-\frac{1}2| D_\mu^{(G)}\eta|^2-\frac{\lambda_2}4\left(|\eta|^2-s^2\right)^2,
\label{1o}\\
\mathcal L_{mix}&=& \frac{\xi}2 G_{\mu\nu}F^{\mu\nu} - \frac{g^2}{\Lambda_g^2} G_{\mu\nu} G^{\mu\nu}|\phi|^2-\frac{\lambda_3}2\left(|\phi|^2-v^2\right)\left(|\eta|^2-s^2\right).
\label{loop}
\eqf

The dynamics of our model will then be governed by the total Lagrangian density ${\cal L}$, 
\be
{\cal L} = \mathcal L_{visible} + \mathcal L_{HS}+ \mathcal L_{mix}.
\label{eq:Lag_dimensionless}
\ee 
From the particle physics point of view, the Higgs portal interaction is usually invoked to solve stability issues of the Standard Model. In this context, let us note that corrections from the portal term can indeed stabilise the potential \cite{Buttazzo:2013uya,Bezrukov:2012sa,Alekhin:2012py}. These issues are particularly relevant when considering the inflationary period \cite{Lebedev:2012sy}. It has been also   established that for the model we discuss here with $\eta$ and $\phi$ developing non-vanishing vevs,  the $\phi-\eta$ portal term induces oscillations between the two scalars in our model \cite{Lebedev:2012zw,Elias-Miro:2012eoi}.

 In order to have a potential bounded from below, the following conditions have to be fulfilled,
\be
\lambda_1, \, \lambda_2>0, \,\,\,\ \lambda_1\lambda_2>\lambda_3^2.
\label{dosiete}
\ee
Given the potentials in Lagrangian \eqref{eq:Lag_dimensionless}  both fields acquire vevs  $\langle \phi\rangle=v$ and $\langle \eta\rangle=s$. The corresponding scalar masses  can be obtained from the  non-diagonal mass matrix
\be
\mathcal M_s=2\begin{pmatrix}
\lambda_1 v^2 & \lambda_3 v s\\
\lambda_3 vs & \lambda_2 s^2
\end{pmatrix}.
\ee
From this, the masses of the propagation eigenstates are given by
\be
m_{1,2}^2={\lambda_1 v^2+s^2 \lambda_2} \pm\sqrt{\left(\lambda_1 v^2- s^2 \lambda_2\right)^2+4\lambda_3^2 s^2 v^2}.\label{eq:masses_scalars}
\ee
This result shows that masses positiveness also requires conditions \eqref{dosiete}. We will define the Higgs-like state as the lighter eigenstate and the hidden scalar as the heavier.

\noindent In the context of Higgs physics, the  $\lambda_3$ value is restricted by observations, together with the mixing angle from the oscillations with the hidden scalar \cite{Weinberg:2013kea}. In our model we will consider an Abelian-Higgs model as a visible Lagrangian density and not the actual Standard Model. Therefore, we will impose a weak coupling between the two sectors by choosing $\lambda_3<1$.

The $\lambda_i$ couplings can be written in terms of the masses of the scalars, a more physical approach to set their values. From eqs.~(\ref{eq:masses_scalars}) one gets   
\eqb
\lambda_{1,2}=
\begin{dcases}
\frac{\Delta+({m_1}^2+{m_2}^2)v^2}{4 v^4},\,\, \frac{-\Delta+({m_1}^2+{m_2}^2)v^2}{4 v^2 s^2},\,\,\,\,\,\,\,\, \mbox{if}\,\,\,\,\, m_1>m_2,\\
\frac{-\Delta+({m_1}^2+{m_2}^2)v^2}{4 v^4},\,\, \frac{\Delta+({m_1}^2+{m_2}^2)v^2}{4 v^2 s^2}, \,\,\,\,\,\,\,\, \mbox{if}\,\,\,\,\,m_2>m_1,
\end{dcases}
\eqf
with
\be
\Delta=\sqrt{\left({m_2}^2-{m_1}^2\right)^2v^4-16
{\lambda_3}^2 s^2
   v^6}.
\ee
In terms of these masses,   conditions \eqref{dosiete} become
\be 
|m_1^2-m_2^2|>4{\lambda_3} sv.
\ee
Concerning the  masses of the vector fields,  let us first cancel the KM interaction by means of the following redefinition
\be \label{rotation}
A_\mu\rightarrow A_\mu+\xi G_\mu,\,\,\,\,\,\mbox{and}\,\,\,\, G_\mu\rightarrow \frac{G_\mu}{\sqrt{1-\xi^2+4\Phi^2_g}},
\ee
with the  dimensionless parameter $\Phi_g$ given by $\Phi_g=gv/\Lambda_g $. It is then easy to   see that the KM term induces a mixing between the two vector fields, which translates into interactions of the vector fields with both scalars, $\phi$ and $\eta$. 
A non diagonal  mass matrix for the vector fields emerge that reads
\be
\mathcal
M_V^2=\begin{pmatrix}
e^2 v^2(1-\xi^2 + 4\Phi_g^2) && && e^2\xi v^2 \sqrt{1-\xi^2+4\Phi_g^2}\\
e^2\xi v^2 \sqrt{1-\xi^2+4\Phi_g^2} && && g^2 s^2+e^2\xi^2 v^2
\end{pmatrix},
\ee
leading to the vector fields masses
\eqb
M_{1,2}^2&=&\frac{1}{2} \left(e^2 v^2+g^2 s^2+4e^2v^2 \Phi_g^2  \pm\sqrt{4 e^2 g^2 s^2
   v^2\xi^2+\left(g^2 s^2-e^2 v^2-4e^2v^2 \Phi_g^2\right)^2}\right).
   \label{eq:masses_vectors}
\eqf
Note that for $\Phi_g=\x=0$, one recovers the expected result $m_A^2=e^2v^2$ and $m_G^2=g^2 s^2$. 

In order to find solutions to the field equations associated to  Lagrangian density \eqref{eq:Lag_dimensionless} it  will be convenient to redefine variables and fields 
\bb
r\rightarrow \frac{r}{(e v)}, \,\,\,\,\, A\rightarrow A v, \,\,\,\,\, G\rightarrow G v,\,\,\,\,\, \phi\rightarrow \phi v, \,\,\,\,\, \eta\rightarrow \eta v,
\ee
 so that that associated  (full) Lagrangian reads
\bb
L=\frac{v}e \int d^3 x \,\left(\mathcal{ \tilde L}_{visible}+\mathcal{\tilde L}_{HS}+\mathcal{\tilde L}_{mix} \right),
\ee
with
 \begin{eqnarray}
\mathcal{ \tilde L}_{visible}&=& -\frac14 F_{\mu\nu} F^{\mu\nu} - \frac12|\mathcal D_\mu^{(A)}\phi|^2 - \frac{\kappa_1}4 (|\phi|^2 - 1)^2, \nonumber  \\
\mathcal{ \tilde L_{HS} }&=& -\frac14 G_{\mu\nu} G^{\mu\nu}-\frac{1}{2}| \mathcal D_\mu^{(G)}\eta|^2-\frac{\kappa_2}4\left(|\eta|^2-\left(\frac{s}{v}\right)^2\right)^2.
\nonumber\\
\mathcal{ \tilde L}_{mix}& =& \frac{\xi}2 G_{\mu\nu}F^{\mu\nu} - \Phi_g^2 G_{\mu\nu} G^{\mu\nu}|\phi|^2-\frac{\kappa_3}2\left(|\phi|^2-1\right)
\left(|\eta|^2-\left(\frac{s}{v}\right)^2\right)
\end{eqnarray}
where  covariant derivatives  read $\mathcal D_\mu^{(A)}=\partial_\mu-i A_\mu$, $\mathcal D_\mu^{(G)}=\partial_\mu-ig_e G_\mu$, where $g_e= g/e$, $\k_i=\lambda_i/e^2$.
 
 Static finite energy solutions of the field equations require  
 $A_0=G_0=0$ \cite{Manias:1985nr}. Moreover,  since we are interested in finding  vortex solutions we will  look for   $z$-independent field configurations proposing the well-honored  Nielsen-Olesen \cite{Nielsen:1973cs}  axially symmetric ansatz for fields in both sectors
\eqb
&&\phi=f(r)e^{in\varphi}, \,\,\, A_\varphi=n\frac{\alpha(r)}r, \,\, A_r=A_z=0, \,\,\, n\in \Z, \label{eq:ansatz_1}\\
&&\eta=h(r)e^{ik\varphi}, \,\,\, G_\varphi=k\frac{ \beta(r)}{g_e r}, \,\, G_r=G_z=0, \,\,\, k\in \Z, \label{eq:ansatz_2}
\eqf 
where 
finite energy solutions require the following behaviour for the functions 
\eqb 
\alpha(0)=\beta(0)=0,\,\,\,\, \f(0)=h(0)=0, 
\alpha(\infty)=\beta(\infty)=1,\,\,\,\mbox{and}\,\,\,\, f(\infty)=1, \,\, h(\infty)= s/v.  
\eqf
The Euler-Lagrange equations read
\begin{eqnarray} 
&&nr\frac{d}{dr}\left(\frac{\a'}r\right)-\frac{\x}{g_e} k r\frac{d}{dr} \left(\frac{\b'}r\right)+n f^2 (1-\a)=0, \label{alpha1ape} \\
&&\frac{k}{g_e} r\frac{d}{dr}\left(\frac{\b'}r\left(1+4\Phi_g^2 f^2\right)\right) - \x n\, r\frac{d}{dr}\left(\frac{\a'}r\right)
+ g_e k h^2 (1-\b)=0,  \\
&& \frac{1}{r}\frac{d}{d\,r}\left(r\,f'\right)-\frac{n^2\,f}{r^2}(1-\a)^2  -4\frac{\Phi_g^2}{g_e^2}
 \frac{k^2}{r^2}\left(\dr\b \right)^2 f - \k_1(f^2-1)f -\k_3 (h^2-s^2/v^2)f  =0,\\
&&\frac{1}{r}\frac{d}{d\,r}\left(r\,h'\right)-\frac{k^2}{r^2}(1-\b)^2h-\k_2(h^2-s^2/v^2)h -\k_3 (f^2-1)h=0. \label{h1ape}
\end{eqnarray}
In order to solve the above system, we implemented a shooting method and also studied the behaviour of the fields  near the origin and infinity (see  Appendix \ref{app:two_unbroken}). In the following we highlight the main results.

Since near  the origin  finite energy solution imply   very small amplitude,  we can neglect non-linear terms in the differential equations. As shown in Appendix \ref{app:two_unbroken} one finds  that  the scalar fields radial functions for  $r\ll 1$ behaves according to
\eqb
f(r)&\approx& r^n\sqrt{\frac{-4 B_{HS}^2 k^2 \Phi_g ^2}{g_e^2}+\kappa
   _1+\frac{\kappa _3 s^2}{v^2}}, \label{eq:f_nearzero}\\
h(r)&\approx& r^k\sqrt{\kappa _3 +\frac{\kappa _2
   }{g_e^2 }\frac{s^2}{v^2}},\label{eq:h_nearzero}
\eqf
Concerning the the  vector fields behavior near the origin one has
\eqb
\alpha(r)\approx c_\alpha r^2,\,\,\,\,\, \beta(r)\approx c_\b r^2,
\eqf
with $c_i$ integration constants. So that the magnetic fields are constant at the origin with values depending on the terms $c_\alpha,c_\beta$ respectively, which in turn    depend on our model parameters. 
 
As for  large distances, both scalar fields acquire vevs, so that the symmetries in the visible and hidden sectors are spontaneously broken. Expanding the fields at large $r$ one can then  solve the resulting   perturbations. Vector fields acquire also a mass and the kinetic mixing explicitly mixes the two vectors. The detailed analysis of the solutions can be found in the Appendix \ref{app:two_unbroken}. For the scalar fields, we find
\eqb
f(r)&\approx& 1+\kappa_{3\rm eff}  \left(\mathcal B_1^+\frac{1}{\sqrt{\tilde m_1 r}} e^{-\tilde m_1r}+\mathcal B_2^-\frac{1}{\sqrt{\tilde m_2 r}} e^{-\tilde m_2r}\right)\\
h(r)&\approx& s/v+ \kappa_{3\rm eff} \left(\mathcal D_1^+\frac{1}{\sqrt{\tilde m_1 r}} e^{-\tilde m_1r}+\mathcal D_2^-\frac{1}{\sqrt{\tilde m_2 r}} e^{-\tilde m_2r}\right),
\eqf
{{ where $\mathcal B_i^{\pm}, \mathcal D_i^{\pm}$ are constants that depend on all the parameters of our model and $\tilde m_i$ are the (dimensionless) expressions for the masses of the scalar fields, given in eq.~(\ref{eq:masses_scalars}), such that $\tilde m_i=m_i/ev$. The  mixing between the two scalars is controlled by the mixing angle, $\kappa_{3\rm eff}$, defined by
\bb
\kappa_{3\rm eff}=\frac{\kappa_3\, s/v }{|\tilde m_1^2-\tilde m_2^2|}.
\ee
As expected,   the large distance fall-off depends on both masses. Mixing also takes place  for the gauge fields because of   the presence of the KM portal. Concerning   functions $\alpha,\beta$ the behave at long distances according to
\bb
\alpha(r), \beta(r) \propto 1+\xi_{\rm eff}\left(\mathcal E_1^{+} \sqrt{r} \,e^{-\tilde M_1 r}+\mathcal E_2^{-} \sqrt{r}\, e^{-\tilde M_2 r}\right),
\ee
with   the vector fields mixing controlled by
\be
\xi_{\rm eff}=\frac{\xi}{|\tilde M_1^2-\tilde M_2^2|}.
\ee
Here $\tilde M_i=M_i/ev$ are given by eq.~(\ref{eq:masses_vectors}) and $\mathcal E^{\pm}_i$ are integration constants.

From the result above we see that  the two characteristic lengths are $r_\phi \approx m_\phi^{-1}$
and $r_\eta \approx m_\eta^{-1}$ indicating the values at which the fields reach their asymptotic values. Concerning the vortex core, their characteristic lengths are related to the inverse mass eigenstates $M_1^{-1}$ and $M_2^{-1}$}}.

\subsection{Numerical results}{\label{subsec1}
As stated above,  we have solved numerically eqs.~\eqref{alpha1ape}-\eqref{h1ape} by using a shooting method finding  an excellent agreement with the limiting results presented in the literature. (Note that in  the absence of portals, parametrized by $\xi$, $\kappa_3$ and $\Phi_g$, the system becomes  two uncoupled copies of a Maxwell-Higgs theory, whose solution is well documented in the literature \cite{Nielsen:1973cs,vilenkin1994cosmic, Hill:1987qx}). Since the KM and HP have been extensively studied before (e.g.~\cite{Hartmann:2009ki, arias2014vortex,Arias:2014yua}), here  we focus on highlighting the behaviour of our model when the SGI portal is present and compare its effect with those of the other two portals. Therefore, the benchmark values for $\kappa_3$ and $\xi$ will be chosen $\sim 0.5$. On the other hand, the benchmark values for $\Phi_g$ will be  considered in the range of $0.5$ and 1. The lower value could be obtained by considering a scale $\Lambda$ in the GeV range and a hidden charge $g>1$. The upper value $\Phi_g=1$ is taken as an extreme case. 
\subsubsection*{SGI-HP portals}

In fig.~\ref{fig:portal_k3P} we present the profile of the scalar and magnetic fields as a function of $r$. The solid gray line is the behaviour of the field in absence of any portal (thus, $\kappa_3=\xi=\Phi_g=0)$, the solid black line turns on only  $\kappa_3$, keeping the other two portals off, and the dashed and dotted lines depicts the system with both HP and SGI portals turned on.  From fig.~\ref{fig:s2_k3P_f} visible scalar field $f(r)$ shows a reduced correlation length when $\kappa_3$ is on, compared to the case where all portals are off. The action of the SGI portal parametrized by $\Phi_g\neq 0$ (dashed and dotted lines) is to counteract on this, forcing $f$ to have a higher correlation length. This effect can be seen already at the linear level of the solution of the fields  from the correlation length at small distances, eq.~(\ref{eq:f_nearzero}). The results can be understood because the SGI portal contributes to the magnetic energy of the hidden sector, at expenses of the energy of the visible sector. Thus, the presence of the SGI portal shall on the one hand delay the scalar field to reach its vev, and on the other hand, decrease the magnetic energy of the visible sector. For the hidden scalar fig.~\ref{fig:s2_k3P_h} we can see that on the one hand the effect of the HP also works shortening the correlation length, but the effect of the SGI portal is to decrease even more the correlation length, as expected. Thus, the energy on the hidden scalar gets higher. 
The behaviour of the magnetic fields follows what has been previously discussed, meaning the visible magnetic field gets diminished in amplitude with the SGI term, while the hidden one gets increased. We also point out that in this case - where $\xi=0$ - the vector fields do not mix, therefore, the fields have definite masses which set the correlation length of the fields. In fig.~\ref{fig:s2_k3P_bv} the mass of the vector field with $\Phi_g=0$ changes slightly than in the case with $\Phi_g\neq0$, and in the latter the eigenvalues are all the same for the benchmark values considered. On the other hand, in fig.~\ref{fig:s2_k3P_bh} the mass eigenvalue for the hidden vector field changes drastically with the change of $\Phi_g$, as can be directly checked from eq.~(\ref{eq:masses_vectors}) and this is reflected in the different correlation lengths for all four curves. 
\begin{figure}[H]
     \centering
     \begin{subfigure}[b]{0.45\textwidth}
         \centering
         \includegraphics[width=\textwidth]{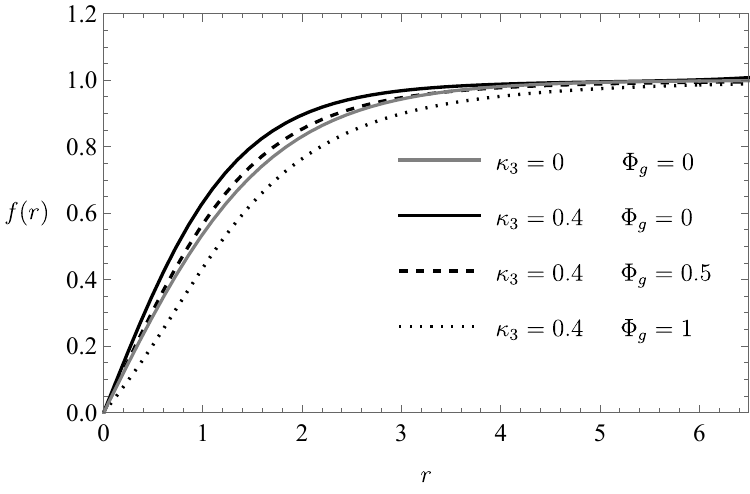}
         \caption{}
         \label{fig:s2_k3P_f}
     \end{subfigure}
     \hfill
     \begin{subfigure}[b]{0.45\textwidth}
         \centering
         \includegraphics[width=\textwidth]{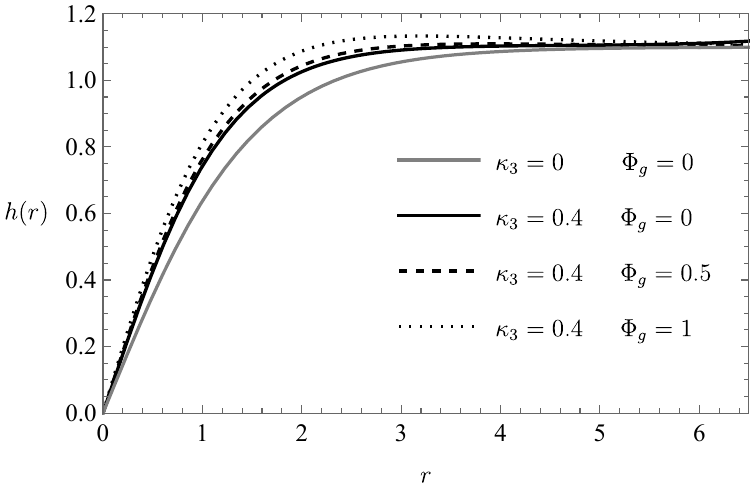}
         \caption{}
         \label{fig:s2_k3P_h}
     \end{subfigure}
     
\bigskip
     \begin{subfigure}[b]{0.45\textwidth}
         \centering
         \includegraphics[width=\textwidth]{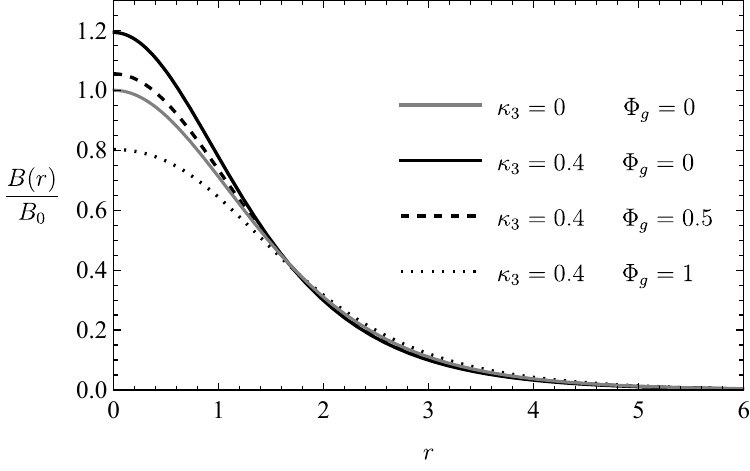}
         \caption{}
         \label{fig:s2_k3P_bv}
     \end{subfigure}
    \hfill
     \begin{subfigure}[b]{0.45\textwidth}
         \centering
         \includegraphics[width=\textwidth]{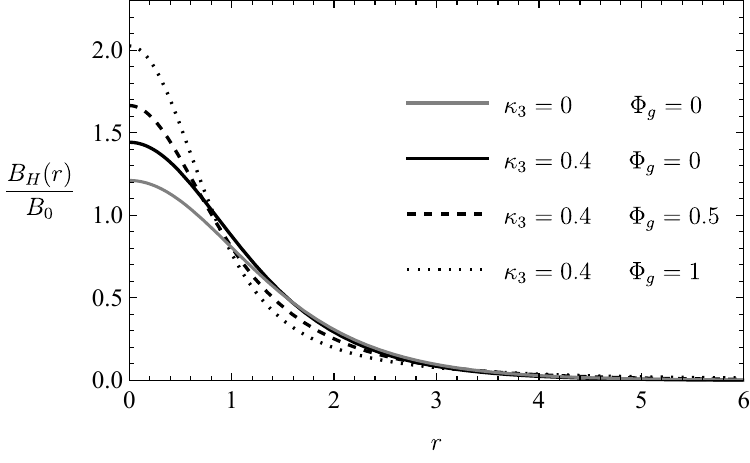}
         \caption{}
         \label{fig:s2_k3P_bh}
     \end{subfigure}
\caption{Effect of the parameters $\kappa_3$ and $\Phi_g$ on the field profiles. (a) Visible scalar field profiles. (b) Hidden scalar field profiles. (c) Visible magnetic field profiles. (d) Hidden magnetic field profiles. The rest of the parameters have been fixed as $\xi=0$, $\kappa_1=\kappa_2=0.5 $ and $s/v$=1.1 } \label{fig:portal_k3P}   
\end{figure}


\subsubsection*{KM-SGI portals}
In fig.~\ref{fig:s2_XP_portal} we study the behaviour of the system when the KM portal (controlled by $\xi$) and SGI portal (controlled by $\Phi_g$) are on, and the rest is off. As in the previous figure, the gray solid line corresponds to the case with all portals off, for comparison. Then, the solid black line corresponds to the case where only the KM parameter is turned on. For this case, it can be seen that the interaction between the gauge fields has a very mild impact on the scalar fields. From figs.~\ref{fig:s2_XP_f} and \ref{fig:s2_XP_h} we can see that the KM itself works augmenting the kinetic energy of the scalar fields, due to the increased magnetic energy at the origin. Asymptotically, though there are not much changes, as the vector fields tend to zero. Once the SGI portal is on (dashed and dotted lines) we can see that it mainly affect the visible scalar, counteracting the action of the kinetic mixing. The reason is that since the scalar field invests some of its energy to increase the hidden magnetic  energy, decreases its kinetic energy. On the contrary, the hidden scalar does not get affected.  Concerning the magnetic fields the action of the KM is to augment their amplitude at the origin, and - as expected - the action of the SGI portal is to increase the hidden magnetic energy, while for the visible magnetic field there is no significant change, except for a change in the mass of eigenstates. But since the magnetic fields considered are composed of a mixture of the eigenstates, the effect in the correlation length is not well defined. We have checked numerically that in order to find significant changes in the visible magnetic field, a SGI term of $\Phi_g \gg 1$ is needed.
\begin{figure}[H]
     \centering
     \begin{subfigure}[b]{0.45\textwidth}
         \centering
         \includegraphics[width=\textwidth]{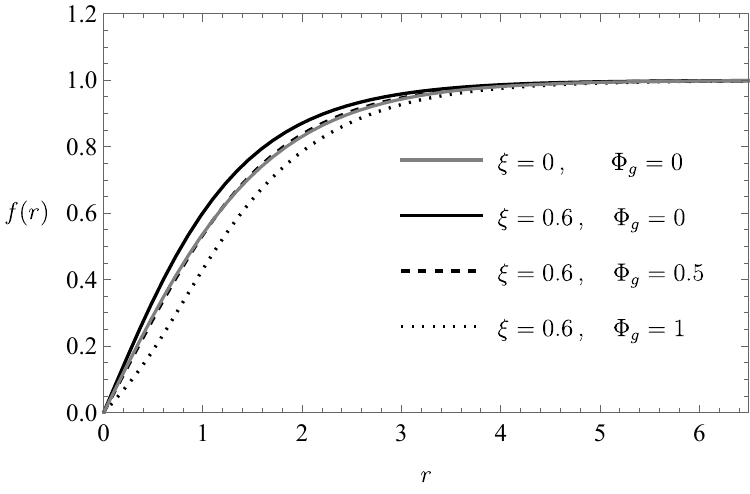}
         \caption{}
         \label{fig:s2_XP_f}
     \end{subfigure}
     \hfill
     \begin{subfigure}[b]{0.45\textwidth}
         \centering
         \includegraphics[width=\textwidth]{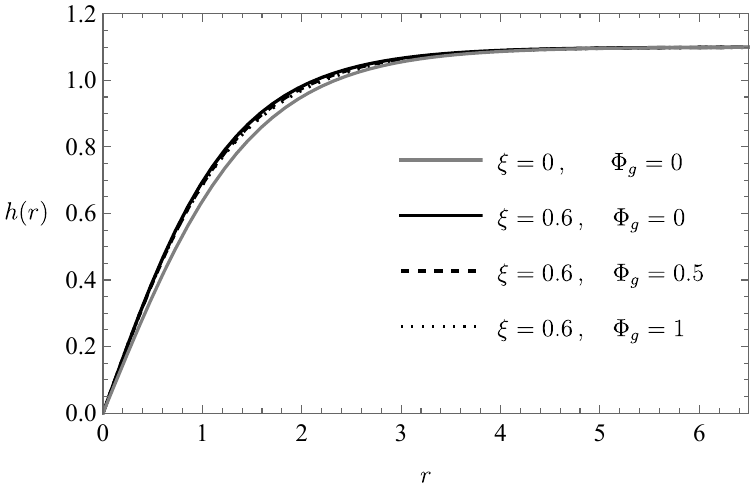}
         \caption{} \label{fig:s2_XP_h}
     \end{subfigure}
     
\bigskip
     \begin{subfigure}[b]{0.45\textwidth}
         \centering
         \includegraphics[width=\textwidth]{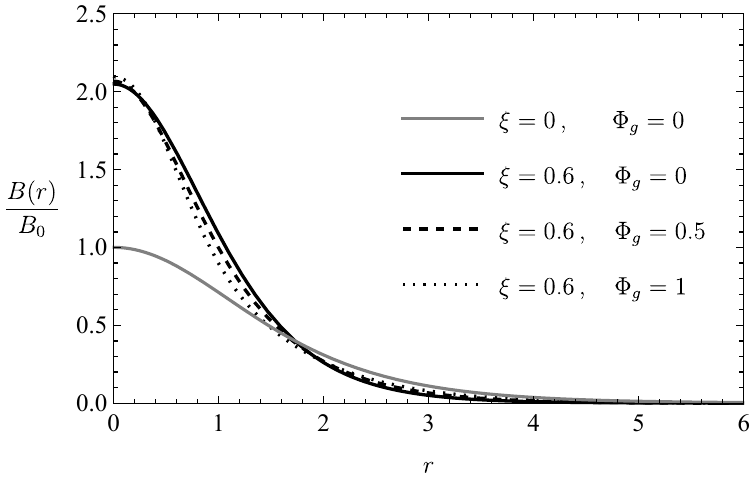}
         \caption{} \label{fig:s2_XP_bv}
     \end{subfigure}
    \hfill
     \begin{subfigure}[b]{0.45\textwidth}
         \centering
         \includegraphics[width=\textwidth]{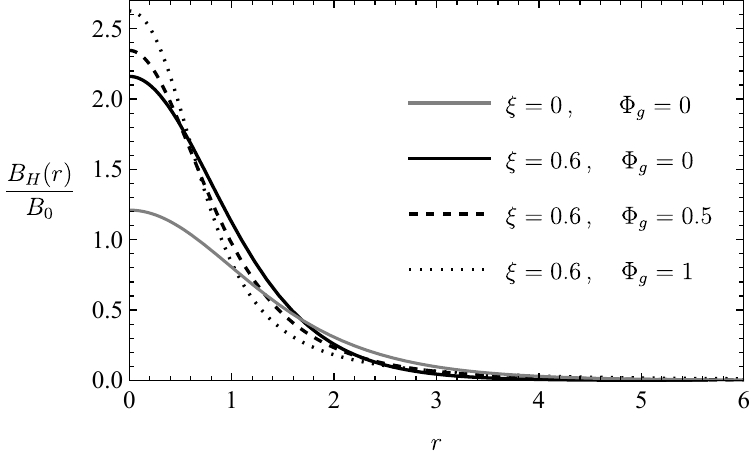} 
         \caption{} \label{fig:s2_XP_bh}
     \end{subfigure}
\caption{Effect of the parameters $\xi$ and $\Phi_g$ on the field profiles. (a) Visible scalar field profiles. (b) Hidden scalar field profiles. (c) Visible magnetic field profiles. (d) Hidden magnetic field profiles. The rest of the parameters have been fixed as $\kappa_3=0$, $\kappa_1=\kappa_2=0.5 $ and $s/v$=1.1 } 
\label{fig:s2_XP_portal}
\end{figure}
\subsubsection*{The action of all portals combined}
In fig.~\ref{fig:s2_XPk3} we show the effect on the model when all portals are active. As before, the gray line corresponds to the case of no-portals, for comparison. The solid line corresponds to the case where the HP and KM portals are turned on, but SGI off. The dashed and dotted lines correspond to the all portals turned on. From fig.~\ref{fig:s2_XPk3_f} we can see that the HP and KM portals have the overall effect of increasing the kinetic energy of the visible scalar, reducing the correlation length. The effect of $\Phi_g$ is to counteract on this effect. The hidden scalar gets some of the benefit, increasing the kinetic energy when $\Phi_g$ increases, as it can be seen in fig.~\ref{fig:s2_XPk3_h}. This is correlated with the profile of the hidden magnetic field  fig.~\ref{fig:s2_XPk3_bh}, that increases its core amplitude due to the three portals, shrinking the correlation length accordingly. Finally, the visible magnetic field, depicted in fig.~\ref{fig:s2_XPk3_bv} increases the amplitude of the core with the HP and KM portal, but it is quite insensitive to the change of $\Phi_g$. From the previous analyses, we have seen that it is the KM the one that shadows the effect of the SGI portal.  Finally, concerning the hidden magnetic field, we see from fig.~\ref{fig:s2_XPk3_bh} that all portals end up enhancing its value near the origin.

\begin{figure}[H]
     \centering
     \begin{subfigure}[b]{0.45\textwidth}
         \centering
         \includegraphics[width=\textwidth]{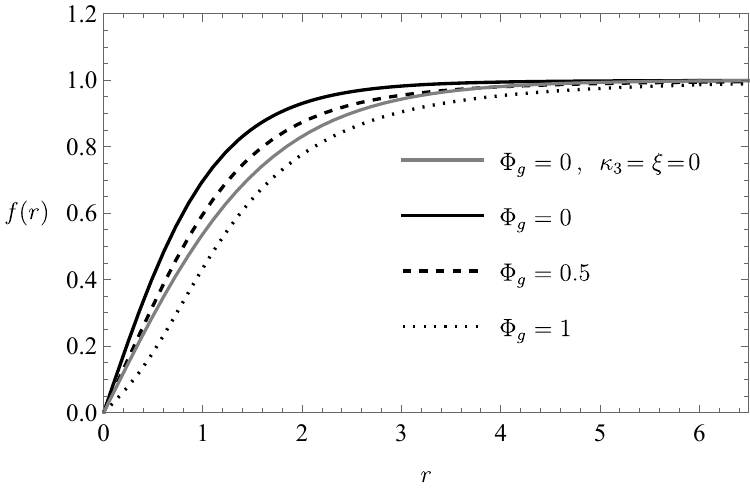}
         \caption{}
         \label{fig:s2_XPk3_f}
     \end{subfigure}
     \hfill
     \begin{subfigure}[b]{0.45\textwidth}
         \centering
         \includegraphics[width=\textwidth]{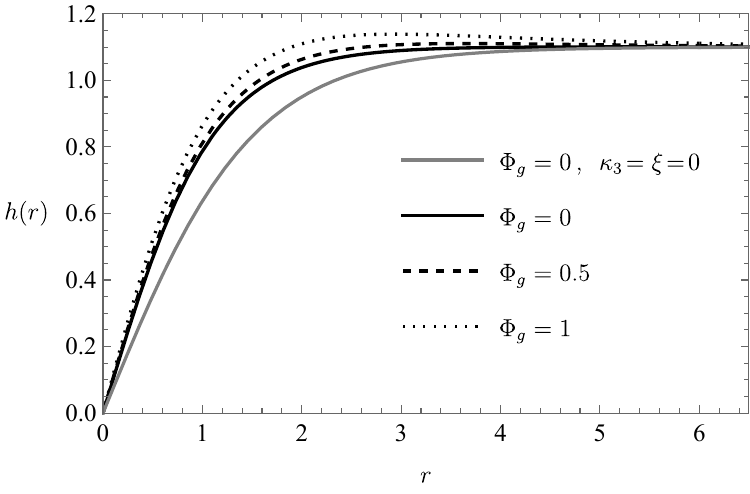}
         \caption{} \label{fig:s2_XPk3_h}
     \end{subfigure}
     
\bigskip
     \begin{subfigure}[b]{0.45\textwidth}
         \centering
         \includegraphics[width=\textwidth]{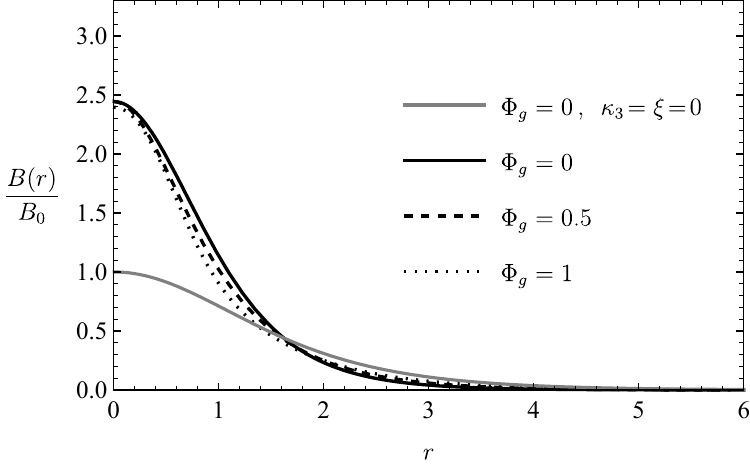}
         \caption{} \label{fig:s2_XPk3_bv}
     \end{subfigure}
    \hfill
     \begin{subfigure}[b]{0.45\textwidth}
         \centering
         \includegraphics[width=\textwidth]{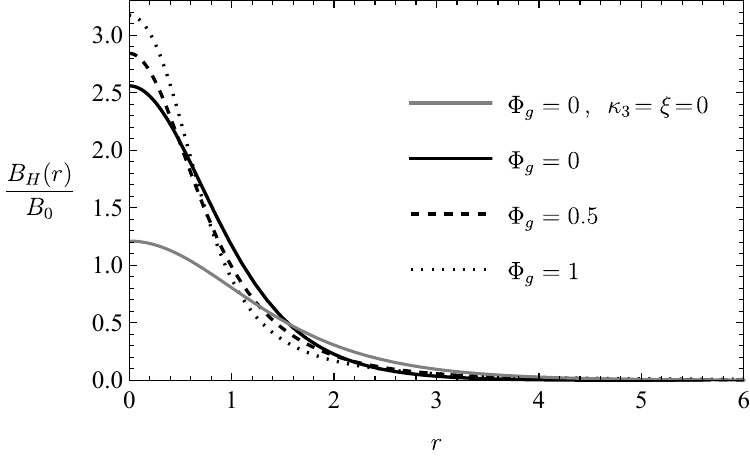}
         \caption{} \label{fig:s2_XPk3_bh}
     \end{subfigure}
\caption{Effect of the parameters $\kappa_3$, $\xi$ and $\Phi$ on the field profiles. (a) Visible scalar field profiles; (b) hidden scalar field profiles; (c) visible magnetic field profiles and (d) hidden magnetic field profiles. The grey line corresponds  to the case where all three portals are turned off, $\kappa_3=\xi=\Phi_g=0$, while for the solid, dotted and dashed black line we set $\kappa_3=0.4$ and $\xi=0.6$. The rest of the parameters have been fixed as $\kappa_1=\kappa_2=0.5 $ and $s/v$=1.1 }  
\label{fig:s2_XPk3}
\end{figure}

\section{Summary and discussion}{\label{sec:conclus}
{We have analyzed in this work vortex solutions of  Abelian-Higgs models describing  both visible and hidden sectors coupled by  a kinetic mixing, a Higgs portal and also a scalar-gauge effective interaction.}

{In section \ref{sectionunbroken} we studied a model in which the visible sector is an  Abelian Higgs with spontaneously broken gauge symmetry while the hidden sector corresponds to an uncharged scalar and a hidden gauge field. Interestingly enough, that kind of models have been studied in connection with different contexts, both in high energy and in condensed matter physics. 
For the hidden scalar, we considered a self-interacting potential and a HP coupling to the visible scalar, depending on three  parameters ($\g,\tilde{\m},\z$) which, if appropriately chosen lead {to expectation values ($\langle \phi\rangle=v, \langle \eta\rangle=0$) at $r \rightarrow \infty$.}}
After the usual gauge field shift   the kinetic mixing portal can be eliminated and the central role is played by the SGI  {and HP portals}. 

The visible system supports  vortex solution. As for the hidden scalar, after a radial dependent ansatz, we found that it develops a non-vanishing value near the origin forming a condensate that radically  changes the physics of the model. Moreover, whenever the condensate emerges, both the energy associated to the  visible scalar and magnetic fields decrease. The parameters that impact the most on the strength of the condensate are the HP portal $\gamma$ and the coupling  , $\gamma \tilde \mu^2$. Interestingly, the effect of the SGI portal is to convey the  the condensate energy to the visible sector,  see fig. \ref{fig:s3_f} and \ref{fig:BUB}. We also showed that for appropriate combinations   of the hidden sector parameters  there is a threshold $\Phi_e$,  where the condensate ceases to exist, as seen in fig.~\ref{fig:s3_plotsh0}.

{In section \ref{broken} we studied a model with two copies of the spontaneously broken Abelian-Higgs model connected by the three mentioned portals. We started by diagonalizing the system, since the HP and KM portals induce oscillations among the scalars and among the vector fields, respectively. Therefore, the interacting fields we are interested in, end up being a superposition of the mass states. After an axially symmetric ansatz we first studied the field behavior at short and large distance and then  solved numerically the field equations applying a shooting algorithm. We found that in order to have consistent solutions near the origin, new relationships between the parameters emerge, besides the ones that are needed in order to have spontaneously symmetry breaking in both sectors. We studied the effect on the vortex solutions for each portal separately (figs.\ref{fig:portal_k3P} and \ref{fig:s2_XP_portal}) and then, analyzed the combined effect in fig.~(\ref{fig:s2_XPk3}). We found that the SGI interaction works increasing the energy of the hidden magnetic field, at expenses of decreasing the energy of the visible scalar. Thus, it counteracts the action of the HP and KM portals. Concerning the hidden scalar behaviour, we have seen KM has little impact on it. On the other hand the SGI interaction $\Phi_g$ has its main impact near the origin, reducing the correlation length, thus, contributing to the kinetic energy of the field. For the visible magnetic field there is a similar effect than for the visible scalar: the SGI portal counteracts on the effect of the HP portal. The latter works increasing the strength of the core, and the SGI reduces it. On the other hand, the effect of the KM shadows the effect of the SGI interaction. For the hidden magnetic field, the SGI portal contributes to its energy, so its action is to increase the strength of the core.

The analytic study of the field behavior near the origin and asymptotically  is presented in Appendices \ref{app:one_unbroken}, \ref{app:two_unbroken}  for each section, respectively.

 Concerning the model discussed in  section \ref{sectionunbroken}, we have shown that  a hidden scalar condensate emerges  for an   appropriate    parameters choice due to an energy transfer, via the Higgs portal, from the visible to the hidden sector. This is a remarkable particle production that does not relies on the thermal mechanisms of dark matter production. It is therefore worth  analyzing in detail models with a hidden sector in connection with the dark matter relic density. The phenomenology of such model is beyond the scope of this work. We hope to come back to this issue in a forthcoming work. 

The results described in section \ref{broken} have  interesting implications concerning  cosmic string network formation and the consequent particle and gravitational waves radiation. The energy  in the hidden vortex  core    increases  due to the SGI interaction, which could be a  relevant feature for  gravitational wave emission but more significantly, for  particle production whenever string segments with dimension
comparable to their thickness, such as  cusps and kinks emerge.  Due to their coupling to the visible sector hidden strings can also induce particle radiation of   the visible sector \cite{Long:2014mxa}.

\section*{Acknowledgments}
{We   thank Fidel I. Schaposnik Massolo for his most valuable help in improving and extending the numerical code.}
D.V. thanks ANID for the support through \textit{Beca Doctorado Nacional $N^o$ 21212328.} {P.A and M.V. acknowledge support from DICYT project, 042131AR-AYUDANTE, VRIDEI. PA acknowledges support from DICYT project 042131AR. F.A.S. is financially supported by grants PIP 0688 and UNLP X850.}

\newpage
\begin{appendix}

\section{Appendix: analytical expressions for the model of section 2} 
\label{app:one_unbroken}
In order to solve numerically the system \eqref{eq:aunbroken}-\eqref{eq:hunbroken} by means of the shooting method, we first find the analytical solutions for both the origin and asymptotic regions, then using our numerical code, we match these localized solutions in between.  

\subsection{Limit $r\rightarrow0$}

Approximated solutions near the origin can be obtained expanding $ h \rightarrow h_r + h_0 $, with $h_r \ll1$ and $ h_0 $ the value of hidden scalar near zero. We recall that $ h_0 $ has a no-vanishing value in this region. Moreover we consider that $f$ and $\a$ are small, nonetheless we asset that the associated magnetic field $B(r)=\a'/r$ is not necessarily negligible. Writing the equation of motion for $\a$ \eqref{eq:aunbroken}, at lower order in this  limit, one gets
\bb
\frac{d}{dr}\left[\left(1+4\Phi_e^2h_0^2 +8\Phi_e^2h_0 h_r\right)B(r)\right]=0.
\ee
From this equation, one easily get a solution for $B(r)$ of the form
\bb
B(r)=\frac{C}{1+\l^2 h_r}, \label{B}
\ee 
where $\l^2\equiv 8\Phi_e^2/(1+4\Phi^2_e\,h_0^2)$ and $C$ as an integration constant. Since $h_r\ll1$, we consider the expansion 
\bb 
\left( \frac{1}{1+\l^2\,h_r}\right)^2\simeq 1-2\l^2h_r + \mathcal{O}(h_r^2), \label{aproximationappendix}
\ee and now we focus on scalars. The scalar fields equations \eqref{eq:funbroken} and \eqref{eq:hunbroken} in this regime can be written as
\bb
\frac{1}{r}\dr (r f')+\left(U_f^2-\frac{n^2}{r^2}\right)f=0,
\ee
\bb
\frac{1}{r}\dr(r h_r')+U_h^2 h_r=-Q\,h_0 , \label{haprox}
\ee
where we collected the parameters in the following definitions:
\eqb
&U_f^2=\k-2\Gamma h_0^2
\\
&U_h^2=2\Gamma\m^2-12\Gamma\z h_0^2-4n^2\Phi_e^2C+8n^2\Phi^2_e\l^2 Ch_0
\\
&Q=2\Gamma \m^2-4\Gamma \z h_0^2-4n^2\Phi_e^2C^2.
\eqf
Note that we arrived at $h$ equation \eqref{haprox} applying the expansion \eqref{aproximationappendix}. Considering the solutions of the equations of motion for the scalars around zero, we finally obtain
\bb
f(r)=f_1 J_n(U_f\,r),
\ee
\bb
h(r)=h_0-\frac{Q\,h_0}{U_h^2}\left(1-J_0(U_h\,r)\right)\approx h_0\left(1-\frac{Q}{4}r^2\right),
\ee
where we have considered the approximation $1-J_0(U_h\,r)\sim \frac{U_h^2}{4}r^2$ around the origin. These solutions are compatible with boundary conditions which we have considered, $f(0)=0$ and $h_r(0)=0$. Concerning to vector field, we re-write \eqref{B} by using $h_r$
\bb 
B(r)
\approx\frac{C}{1+\frac{\l^2 Q}{4} r^2}.
\ee 
\newline By completeness, we look for an expression to $\a$. Integrating the last expression
\bb 
\a(r)=C\int \frac{r}{1+\frac{\l^2 Q}{4} r^2}\,dr
\ee 
and imposing the condition $\a(0)=0$, the first terms of an approximate solution around $r\sim 0$ are
\bb
\a(r)\simeq \frac{C}{2}r^2-\frac{C}{16}Q\l^2\,r^4+\mathcal{O}(r^6).
\ee 

\subsection{Limit $r\rightarrow\infty$}

We start assuming that all the fields converge to a constant value at $r\rightarrow\infty$ given by their corresponding vacuum expectation value.
From equation \eqref{eq:aunbroken} we can see immediately that $\alpha\rightarrow1$. For $f$ and $h$, we keep considering the discussion according to section \ref{sectionunbroken}, then asymptotically the fields turn $f \rightarrow 1$ and $h\rightarrow 0$ in order to reach a stable minimum. Replacing with the new variables $\a \sim 1- \tilde{\a}$ and $f \sim 1 -\tilde{f}$, at first order the system can be cast as 
\begin{eqnarray}
\dr\left(\frac{\tilde{\a}'}{r}\right)-n\frac{\tilde{\a}}{r}=0,
\\
\frac{1}{r}\dr (r \tilde{f}')-\frac{m_\f^2}{e^2v^2}\tilde{f}=0, \\
\frac{1}{r}\dr (r h')-\frac{m_\h^2}{e^2v^2}h=0.
\end{eqnarray} 
Imposing the condition that the fields converge, the general solutions for them, in terms of modified Bessel function $K_n$, are
\eqb 
&&\a(r)=1+\a_1\,r K_1(r), \\ 
&& f(r)=1+f_1 K_0\left(\frac{m_\f}{e v}r\right), \\
&& h(r)=h_1 K_0\left(\frac{m_\h}{e v}r\right).
\eqf


\section{Appendix: analytical expressions for the model of section 3}{\label{app:two_unbroken}}
In the same way as the previous section we will apply the shooting algorithm, now to solve the system \eqref{alpha1ape}-\eqref{h1ape}. 
\subsection{Limit $r\rightarrow 0$}
Near the origin, the fields amplitude are expected to be small, so the system of equations at the first order in fields become 
\eqb
-k\frac{\xi  \beta ''(r)}{g_e}+k\frac{\xi  \beta '(r)}{g_e r}+n \alpha
   ''(r)-n\frac{\alpha '(r)}{r}&=&0,\\
k\frac{\beta ''(r)}{g_e}-k\frac{\beta '(r)}{g_e r}-{n\xi 
   \alpha ''(r)}+n\frac{\xi  \alpha '(r)}{ r}&=&0,\\
   f''(r)+\frac{f'(r)}{r}+
   \left(-\frac{4 \Phi_g^2 k^2}{g_e^2}\frac{\beta'(r)^2}{r^2} -\frac{n^2}{r^2}+\kappa_1+\frac{\kappa_3 s^2}{v^2}\right)f(r)&=&0,\\
   h''(r)+\frac{h'(r)}{r}+\left(-\frac{k^2}{r^2}+\kappa_3 +\frac{\kappa_2 s^2}{v^2}\right) h(r)&=&0.
\eqf
The scalar equations of motions can be solved directly. For $f(r)$ the $n$th order turn out to be
\be
f(r)= c_1 J_n\left( r \sqrt{\frac{-4 B_{HS}^2 k^2 \Phi_g ^2}{g_e^2}+\kappa
   _1+\frac{\kappa _3 s^2}{v^2}}\right)\approx r^n\sqrt{\frac{-4 B_{HS}^2 k^2 \Phi_g ^2}{g_e^2}+\kappa
   _1+\frac{\kappa _3 s^2}{v^2}},\,\,\, r\ll1
\ee
where $\beta'/r=B_{HS}$, is the magnetic field of the hidden vector field, which we take near the origin as constant and $J_n(r)$ is the Bessel function of first kind. Similarly, the solution for hidden scalar $h(r)$ near zero is given by 
\be
h(r)=c_2\,J_k\left({r \sqrt{\kappa _3 +\frac{\kappa _2
   }{g_e^2 }\frac{s^2}{v^2}}}\right)\approx r^k\sqrt{\kappa _3 +\frac{\kappa _2
   }{g_e^2 }\frac{s^2}{v^2}},\,\,\,\,\,\mbox{for}\,\,r\ll1.
\ee
To have finite energy solutions near $r=0$, and considering  $\k_i$ to be positive, the above solutions lead to the following  requirements 
\be
\k_1+\k_3\frac{s^2}{v^2}> \frac{4k^2\Phi_g^2}{g_e^2} B_{HS}^2,
\ee

\be
{\kappa _2
 }{g_e^2 }\frac{s^2}{v^2}+\k_3>0.
\ee
On the other hand, the solutions for the functions $\alpha$ and $\beta$ of the vector fields are easily found in this limit to be
\be
\alpha(r)\sim c_\a\,r^2, \,\,\, \beta(r)\sim c_\b\, r^2,\,\,\,\,\,\, \mbox{for}\,\,\,r\ll1,
\ee
with $c_\a$ and $c_\b$ integration constants.
\subsection{Limit $r\rightarrow \infty$}
Concerning the asymptotic behavior, in order to have finite energy vortex solutions all fields should reach their vacuum expectation values. We can again linearize the governing equations by considering a fluctuation around their vevs, as 
\be
f(r)\sim 1-\tilde f(r),\,\,\,\,\, h(r)\sim \frac{s}{v}-\tilde h(r),\,\,\,\,\, \alpha(r)\sim1-\tilde \alpha(r), \,\,\,\,\,\beta(r)\sim 1-\tilde \beta(r).
\ee
Thus, by linearizing the equations in the asymptotic limit, the following system shall be solved 
\eqb
n\,\rdr{\tilde \a'(r)}-\x \frac{k}{g_e}\, \rdr{\tilde \b'(r)}+n\tilde\a(r)&=&0, \\
\frac{k}{g_e}\,\rdr {\tilde \b'(r)} -\x n\,\rdr{\tilde \a'(r)} -4\Phi_g^2\frac{k}{g_e r}\tilde \b'(r)+4\Phi_g\frac{k}{g_e}\tilde \b''(r)+\frac{k}{g_e} \s^2 \tilde \b(r)&=&0, \\
\tilde f''(r)+\frac{\tilde f'(r)}{r}+2\k_1\tilde f(r)
+ 2\k_3 \frac{s}{v}\, \tilde h(r)&=&0, \\
\tilde h''(r)+\frac{\tilde h'(r)}{r}+2\k_2{\frac{s^2}{v^2}}\,\tilde h(r) +2\k_3{\frac{s}{v}}\, \tilde f(r)&=&0.
\eqf
We have defined $\s^2=g_e^2 s^2/v^2$. Although a bit tedious, the system can be solved analytically following a similar procedure used in a previous work \cite{arias2014vortex}. For the vector fields, one finds the expressions 
\eqb
\tilde{\a}=n \frac{\gamma_1(r) - \gamma_2(r)}{\delta_1 \, - \, \delta_2},\,\,\,\,\,\tilde{\b}=\frac{k}{g_e\,\s^2}\frac{\gamma_2(r)\d_1 - \gamma_1(r)\d_2 }{\delta_1- \delta_2 },
\eqf
while the radial functions of the scalar fields are found to be
\eqb
\tilde{f}=\frac{\nu_1(r)-\nu_2(r)}{\epsilon_1 \, - \, \epsilon_2},\,\,\,\,\,\,
\tilde{h}=\frac{\nu_1(r)\epsilon_2-\nu_2(r)\epsilon_1}{\epsilon_2 \, - \, \epsilon_1}.
\eqf
We have defined the terms
\eqb
    \gamma_{1,2}(r)&=&A_{1}^{1,2}\, r I_1\left(\sqrt{C_{1,2} }r\right)+ A_2^{1,2}\,r K_1\left(\sqrt{C_{1,2} }r\right),\\
    \nu_{1,2}(r)&=&B_1^{1,2}\,I_0(\sqrt{D_{1,2}}\,r)+ B_2^{1,2}\,K_0(\sqrt{D_{1,2}}\,r).
\eqf
\bb
C_{1,2}=\frac{M_{1,2}^2}{g^2 s^2},\,\,\,\,\, D_{1,2}=\frac{m^2_{1,2}}{e^2v^2},
\ee
\begin{eqnarray}
\delta_{1,2}=\frac{\s^2-M_{1,2}^2/(e^2v^2)}{\x }.
\end{eqnarray}
\bb
\epsilon_{1,2}=\frac{m^2_{1,2}/e^2v^2-2\k_2s^2/v^2 }{2\k_3s/v}.
\ee

\end{appendix}
\bibliographystyle{unsrt}
\bibliography{bibliog.bib}

\end{document}